\documentclass[12pt]{iopart}
 \expandafter\let\csname equation*\endcsname\relax
  \expandafter\let\csname endequation*\endcsname\relax
\usepackage{amsmath,amsfonts,amssymb,bm,graphicx,mathrsfs,units,cite}

\usepackage{color,colordvi}
\definecolor{blue}{rgb}{0,0,1}
\definecolor{grey}{rgb}{0.6,0.6,0.6}

\definecolor{myurlcolor}{rgb}{0,0,0.7}
\definecolor{myrefcolor}{rgb}{0.8,0,0}
\usepackage[unicode=true,pdfusetitle, bookmarks=false,bookmarksnumbered=false,
bookmarksopen=false, breaklinks=false,pdfborder={0 0 0},backref=false,
colorlinks=true, linkcolor=myrefcolor,citecolor=myurlcolor,urlcolor=myurlcolor]{hyperref}

\newcount\colveccount
\newcommand*\colvec[1]{
        \global\colveccount#1
        \begin{pmatrix}
        \colvecnext
}
\def\colvecnext#1{
        #1
        \global\advance\colveccount-1
        \ifnum\colveccount>0
                \\
                \expandafter\colvecnext
        \else
                \end{pmatrix}
        \fi
}

\newcommand{\figref}[1]{Fig.~\ref{#1}}

\newcommand{\secref}[1]{Sec.~\ref{#1}}

\newcommand{\ket}[1]{\left| #1 \right \rangle}
\newcommand{\bra}[1]{\left \langle #1 \right|}

\begin{document}


\title[Markovian master equations for quantum thermal machines]{Markovian master equations for quantum thermal machines: local vs global approach}

\author{Patrick P. Hofer$^1$, Mart\' i Perarnau-Llobet$^2$, L. David M. Miranda$^1$, G\'eraldine Haack$^1$, Ralph Silva$^1$, Jonatan Bohr Brask$^1$, and Nicolas Brunner$^1$}

\address{$^1$ D\'epartement de Physique Appliqu\'ee, Universit\'e de Gen\`eve, Switzerland}
\address{$^2$ Max-Planck-Institut f\"ur Quantenoptik, Hans-Kopfermann-Str. 1, D-85748 Garching, Germany}

\ead{patrick.hofer@unige.ch}

\date{\today}

\begin{abstract}
The study of quantum thermal machines, and more generally of open quantum systems, often relies on master equations. Two approaches are mainly followed. On the one hand, there is the widely used, but often criticized, local approach, where machine sub-systems locally couple to thermal baths. On the other hand, in the more established global approach, thermal baths couple to global degrees of freedom of the machine. There has been debate as to which of these two conceptually different approaches should be used in situations out of thermal equilibrium. Here we compare the local and global approaches against an exact solution for a particular class of thermal machines. We consider thermodynamically relevant observables, such as heat currents, as well as the quantum state of the machine. Our results show that the use of a local master equation is generally well justified. In particular, for weak inter-system coupling, the local approach agrees with the exact solution, whereas the global approach fails for non-equilibrium situations. For intermediate coupling, the local and the global approach both agree with the exact solution and for strong coupling, the global approach is preferable. These results are backed by detailed derivations of the regimes of validity for the respective approaches.
\end{abstract}


\vspace{2pc}
\maketitle
\section{Introduction}

Master equations are a powerful tool to study open quantum systems \cite{breuer:book,weiss:book}. They allow for a description of the relevant degrees of freedom only, which evolve under the influence of all other degrees of freedom that are not of immediate interest. These other degrees of freedom are collectively called the environment. A particularly simple situation occurs when the system can be described by a time-local master equation with constant dissipation rates \cite{gorini:1976,lindblad:1976,gorini:1978,rivas:book}. This results in Markovian evolution, where knowledge of the density matrix at a given time is sufficient to predict all future observables, which implies an environment that has no memory. Here we refer to this type of master equations as Markovian. 

Compared to the complete problem of describing all the degrees of freedom of system and environment together, a Markovian master equation governing only the system degrees of freedom is an immense simplification.
Such a drastic reduction of complexity usually comes at a price. In this case, the price comes in the form of strong approximations which are not always justified. Studying these approximations is thus of utmost importance and indeed, there is a large body of literature that addresses these issues \cite{davies:1974,davies:1976,spohn:1978,gorini:1978,spohn:1980,harbola:2006,alicki:2006,scala:2007,scala:2007jpa,wichterich:2007,rivas:2010,albash:2012,Thingna2013,Santos2014, manrique:2015,barra:2015,Trushechkin2016,Purkayastha2016,brandner:2016,seja:2016,goldozian:2016,Stockburger2017,DECORDI2017,boyanovsky:2017b} (for a recent review, see Ref.~\cite{rivas:book}). However, a large number of these studies focus on an environment that drives the system towards equilibrium. The recent rise in interest in out-of-equilibrium quantum systems, and in particular in quantum thermodynamics, calls for revisiting the question of the validity of the widely used Markovian quantum master equations \cite{wichterich:2007,rivas:2010,Thingna2013,manrique:2015,barra:2015,Trushechkin2016,Purkayastha2016,brandner:2016,seja:2016,goldozian:2016}.

Quantum thermal machines are devices that perform useful tasks by exploiting thermal gradients in the environment; for recent reviews, see e.g. \cite{kosloff:2014,gelbwaser:2015,goold:2016,vinjanampathy:2016,benenti:2017,kosloff:2013}. This task can for instance be the production of work \cite{kosloff:1984,quan:2007,brunner:2012,abah:2012,roulet:2017,hardal:2017}, or more concretely of an electrical current \cite{sothmann:2015,hofer:2015,hofer:2016prb,benenti:2017}, the refrigeration of a quantum degree of freedom \cite{palao:2001,linden:2010prl,levy:2012,hofer:2016,mitchison:2016,maslennikov:2017}, the creation of entanglement \cite{brask:2015njp,boyanovsky:2017,tavakoli:2017}, the determination of low temperatures \cite{hofer:2017}, or the design of thermal transistors \cite{joulain:2016} and autonomous quantum clocks \cite{erker:2017}. 

The standard description of these systems crucially relies on Markovian master equations to predict the relevant observables, such as heat currents and power. Two main approaches are followed in the literature. The first is a local approach, where the thermal baths couple locally to sub-systems of the machine. The second is a global approach, where thermal baths couple to the global eigenmodes of the machine. As the two approaches are conceptually different, there has been considerable debate about which one should be used in order to accurately describe thermal machines, and more generally out-of-equilibrium systems. Since the global approach describes equilibrium situations accurately (see below), while the local in some cases does not, there has been incentive to use the global approach out of equilibrium as well. Furthermore, the local approach is often believed to be more phenomenological in nature \cite{scala:2007,scala:2007jpa,Migliore:2011,Santos2014,DECORDI2017} and it was even argued that it is unphysical in certain regimes \cite{Capek:2002,levy:2014,Stockburger2017}. 

The goal of the present work is to discuss these questions in depth. We will consider a system for which the full unitary dynamics of the machine and the thermal baths can be solved exactly. This allows us to evaluate the performance of local and global master equations for the machine against the exact dynamics. In addition, we give detailed derivations of the local and the global approaches and discuss the involved approximations. Specifically, we consider a heat engine introduced by Kosloff \cite{kosloff:1984}, which can be implemented in superconducting circuits \cite{hofer:2016prb}. The machine consists of two sub-systems (oscillators), which couple to different thermal baths, and to each other via an energy conserving interaction. In case the two oscillators have different frequencies, the machine requires an external driving field, making the Hamiltonian time-dependent. The entire system (machine plus baths) consists only of harmonic oscillators with quadratic interactions. Therefore, the system can be described exactly at the level of covariance matrices, and can be treated numerically even for relatively large baths with arbitrary precision. This exact numerical solution serves as a benchmark for evaluating the performance of both the local and global master equations. Focusing on an out-of-equilibrium steady-state regime, we discuss relevant thermodynamical observables, such as heat currents and power, as well as the quantum state (density matrix) of the machine degrees of freedom. 

Our results demonstrate an overall excellent agreement between the predictions of the local approach and the exact solution. 
In the weak inter-system coupling regime, we see that the local approach provides an accurate description of the system, capturing both the thermodynamical observables and the quantum state. On the contrary, the global approach fails in this regime. Moving to the regime of intermediate coupling, we find that both approaches provide good descriptions of the system. Notably, the local approach still reliably captures all thermodynamical features of the machine. For very strong inter-system coupling strengths, the local approach starts to fail while the global approach still yields a faithful description of the system. We provide a detailed derivation of the regime of validity for each approach.

In the final part of the paper, we briefly discuss the case of finite dimensional machines. In particular, we consider the two-qubit entangler of Ref.~\cite{brask:2015njp}, which is analogous to the heat engine setup considered in the first part, but with the two machine oscillators replaced by two qubits with equal level spacing (i.e. no external drive). While solving the total system (including the baths) exactly is unfortunately out of reach in this case, we can still compare the local and global approaches. We find very similar behavior to the results of the first part. In particular, the global approach still fails in the weak coupling regime, while for intermediate coupling, the two approaches agree well.

The paper is structured as follows. Section \ref{sec:localglobal} gives a more detailed introduction to the local and global approaches. Sections \ref{sec:heatengine}-\ref{sec:results} are devoted to the heat engine. In Sec.~\ref{sec:heatengine}, we introduce the system. The different master equations and the respective approximations are discussed in Sec.~\ref{sec:mastereqs}, and the exact numerics are discussed in Sec.~\ref{sec:numerics}. The observables which are investigated are introduced in Sec.~\ref{sec:obs} and the results are given in Sec.~\ref{sec:results}. The qubit entangler is then discussed in Sec.~\ref{sec:qubits} before we conclude in Sec.~\ref{sec:conclusions}.

\section{Local vs global}\label{sec:localglobal}

Before going into details, we provide a short introduction to the two commonly used Markovian master equations which we discuss. In the local approach, the thermal baths couple to the eigenstates of sub-systems of the machine, while the global approach is based on a secular approximation. For time-independent Hamiltonians, the global approach corresponds to baths that couple to the delocalized eigenstates of the system Hamiltonian. In an equilibrium situation (i.e. baths at equal temperatures and time-independent Hamiltonian) the global master equation results in the desired steady state which is given by a Gibbs state with respect to the system Hamiltonian \cite{breuer:book,carmichael:1973} (for a discussion on deviations from the Gibbs state due to the finite coupling between system and bath, see, e.g., Ref.~\cite{geva:2000,subasi:2012}). The local approach on the other hand results in a product of Gibbs states with respect to the sub-system Hamiltonians \cite{carmichael:1973}. For finite interactions, the global master equation is therefore usually considered superior to the local master equation (cf.~Fig.~\ref{fig:equilibrium}). The situation drastically changes if we move away from equilibrium. Clearly, if the sub-systems do not interact, each sub-system should thermalize to its respective bath. While the local master equation yields correct results in this case, the global approach fails, resulting in a finite energy current through the system even in the absence of an interaction (cf.~Fig.~\ref{fig:degen_g}) There therefore exist limiting cases where we expect one approach to clearly be superior to the other. The goal of the present work is to connect these dots by comparing the local and the global master equations to exact numerics for a wide range of parameters in and out of equilibrium. Our main interest thus lies in the observables related to the machine operation, i.e. heat currents, powers, and efficiencies. Moreover, we also compare the quantum states of the machine.

We note that the validity of local and global approaches has been investigated before, even for out-of-equilibrium systems. Ref.~\cite{rivas:2010} provides a detailed study of the different approximations, however, the authors only consider the state of the system and not the energy flows. Ref.~\cite{harbola:2006} shows that the global approach neglects terms that influence the current through a two-terminal electric conductor. In Ref.~\cite{wichterich:2007}, it was shown that the global approach can erroneously result in a vanishing heat current through a spin chain while a local approach shows good agreement with results obtained from a Redfield equation. The validity of the Redfield equation was shown in Ref.~\cite{Purkayastha2016}, where a system analogous to our heat engine in the absence of an external drive is investigated. Refs.~\cite{Capek:2002,levy:2014,gonzalez:2017} argue that a local master equation can violate the second law of thermodynamics when a non-energy preserving interaction between the sub-systems is considered. These violations were however shown to be of the order of terms that are dropped when deriving a local master equation \cite{novotny:2002,Trushechkin2016}. In Ref.~\cite{Trushechkin2016}, it was shown that in the absence of degenerate subspaces, the local approach can be understood as the zeroth order of a perturbation series in the inter-system interaction. However, most thermal machines cited above crucially rely on such degenerate subspaces. Finally, Ref.~\cite{manrique:2015} investigates heat currents through a system that is equivalent to our qubit entangler, without providing a benchmark. Their results agree with the results presented in Sec.~\ref{sec:qubits}. All these previous works motivate our detailed investigation which compares the different master equations for a wide range of parameters.

\begin{figure}[h!]
\centering
\includegraphics[width=.75\columnwidth]{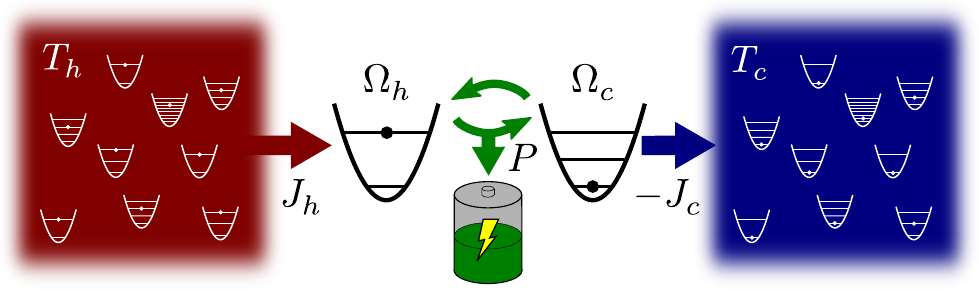}
\caption{Sketch of the heat engine. The system consists of two harmonic oscillators with different frequencies. Each oscillator couples to a thermal bath modeled by a collection of harmonic oscillators. An external field mediates a weak coupling between the two oscillators. A thermal gradient can result in a heat flow from the hot to the cold bath, injecting some of the energy into the external field which can in principle be used to charge a battery. A thermoelectric implementation of this machine in superconducting circuits is proposed in Ref.~\cite{hofer:2016prb}. In this work, we compare different Markovian master equations for the description of this machine, using exact numerics as a benchmark.}
\label{fig:sketch}
\end{figure}

\section{Heat engine}
\label{sec:heatengine}
The heat engine we consider \cite{kosloff:1984,hofer:2016prb} is sketched in Fig.~\ref{fig:sketch} and consists of two harmonic oscillators, with frequencies $\Omega_c$ and $\Omega_h$, which each couple to a bath. Here the subscript labels both the oscillators as well as the corresponding bath (where $c$ stands for cold and $h$ for hot). In this setup, the oscillators constitute the weakly interacting sub-systems mentioned above. The frequencies of the oscillators differ from each other by
\begin{equation}
\label{eq:drive}
\mathcal{E}=\Omega_h-\Omega_c\geq 0.
\end{equation}
In order to extract power from the machine, we consider a time-dependent external field with frequency $\mathcal{E}$ which mediates a coupling between the harmonic oscillators. The purpose of the heat engine is then to use a heat flow from the hot bath to the cold bath in order to increase the power of this external field. In Ref.~\cite{hofer:2016prb}, this external field is provided by a voltage and the power is directly related to an electrical current flowing against the voltage.

The total Hamiltonian of the heat engine (including heat baths) can then be written as (throughout this paper we set $\hbar=1$)
\begin{equation}
\label{eq:hamiltonian}
\begin{aligned}
&\hat{H}^l(t)=\hat{H}^l_S(t)+\sum_{\alpha=c,h}\left[\hat{H}^l_{\alpha}+\hat{V}_{\alpha}\right],\\
&\hat{H}^l_S(t)=\sum_{\alpha=h,c}\Omega_\alpha\hat{a}_\alpha^\dagger\hat{a}_\alpha+g\left(\hat{a}^\dagger_c\hat{a}_he^{i\mathcal{E}t}+H.c.\right),\\
&\hat{H}^l_\alpha=\sum_k\omega_{k,\alpha}\hat{b}_{k,\alpha}^\dagger\hat{b}_{k,\alpha},\hspace{1cm}\hat{V}_\alpha=\sum_k\gamma_{k,\alpha}\left(\hat{b}_{k,\alpha}^\dagger\hat{a}_\alpha+\hat{a}_\alpha^\dagger\hat{b}_{k,\alpha}\right),
\end{aligned}
\end{equation}
where $\hat{a}_\alpha$ ($\hat{b}_{k,\alpha}$) denote annihilation operators of the system (baths), $g$ denotes the interaction strength between the harmonic oscillators of the system, $\omega_{k,\alpha}$ are the frequencies of the bath modes, and $\gamma_{k,\alpha}$ denote the interaction strengths between system and bath modes. The superscript $l$ denotes the laboratory frame. The external drive accounts for the energy that is needed to convert a photon with frequency $\Omega_c$ into a photon with frequency $\Omega_h$ and vice versa. In contrast to Refs.~\cite{Capek:2002,levy:2014,gonzalez:2017}, which lack an external field, we find that the presence of such a field ensures that the local master equation does not result in any violations of the laws of thermodynamics.

In the following, we will work in a rotating frame defined by the transformation
\begin{equation}
\label{eq:unitrot}
\hat{U}_r(t)=\exp\left[it\sum_{\alpha=h,c}\Omega_\alpha\left(\hat{a}^\dagger_\alpha\hat{a}_\alpha+\sum_k\hat{b}^\dagger_{k,\alpha}\hat{b}_{k,\alpha}\right)\right].
\end{equation}
This results in the time-independent Hamiltonian
\begin{equation}
\label{eq:hamiltonianrot}
\begin{aligned}
\hat{H}&=\hat{U}_r(t)\hat{H}^l(t)\hat{U}^\dagger_r(t)-i\hat{U}_r(t)\partial_t\hat{U}^\dagger_r(t)=\hat{H}_S+\sum_{\alpha=c,h}\left[\hat{H}_{\alpha}+\hat{V}_{\alpha}\right],\\
\hat{H}_S&=g\left(\hat{a}^\dagger_c\hat{a}_h+H.c.\right),\hspace{1cm}
\hat{H}_\alpha=\sum_k(\omega_{k,\alpha}-\Omega_\alpha)\hat{b}_{k,\alpha}^\dagger\hat{b}_{k,\alpha}.
\end{aligned}
\end{equation}
Note that the interaction terms $\hat{V}_\alpha$ are invariant under the transformation.

If not explicitly stated otherwise, all equations are given in the rotating frame.

\section{Master equations}
\label{sec:mastereqs}
In this section, we consider different Markovian master equations which are used to describe the evolution of the reduced density matrix of the system. These equations allow for analytic expressions of observables such as power, heat, and efficiency which are then compared to the ones obtained from the exact dynamics.
The standard way of deriving Markovian master equations is to first perform Born-Markov approximations. This procedure is discussed in detail elsewhere (see for instance Refs.~\cite{breuer:book,carmichael:book,rivas:2010,rivas:book}). We therefore only summarize the approximations and give the resulting expression for the system under consideration. The approximations are:
\begin{itemize}
\item Born approximation: Treating $\hat{V}_\alpha$ perturbatively to lowest order,
\item Markov approximation: Assuming invariance of $\tilde{\rho}(t)$ on time-scales of the order of $\tau_B$,
\end{itemize}
where $\tilde{\rho}(t)$ is the reduced density matrix in the interaction picture and $\tau_B$ denotes the bath-correlation time that will be introduced below.
In the interaction picture (and the rotating frame), the Born-Markov approximations result in the following master equation,
\begin{equation}
\label{eq:bornmarkovint}
\begin{aligned}
\partial_t\tilde{\rho}(t)=\sum_{\alpha=h,c}&\int_{0}^{\infty}d\tau\bigg\{C_{1,\alpha}(\tau)\left[\tilde{a}^\dagger_\alpha(t-\tau)\tilde{\rho}(t)\tilde{a}_\alpha(t)-\tilde{a}_\alpha(t)\tilde{a}^\dagger_\alpha(t-\tau)\tilde{\rho}(t)\right]\\&
+C_{2,\alpha}(\tau)\left[\tilde{a}_\alpha(t-\tau)\tilde{\rho}(t)\tilde{a}^\dagger_\alpha(t)-\tilde{a}^\dagger_\alpha(t)\tilde{a}_\alpha(t-\tau)\tilde{\rho}(t)\right]\bigg\}+H.c.
\end{aligned}
\end{equation}
Here operators in the interaction picture are given by
\begin{equation}
\tilde{A}(t)=\hat{U}^\dagger_S(t)\hat{A}\hat{U}_S(t),
\end{equation}
where $\hat{A}$ denotes the operator in the Schr\"odinger picture and
\begin{equation}
\label{eq:unitsys}
\hat{U}_S(t)=e^{-i\hat{H}_St}.
\end{equation}
We further introduced the bath correlation functions
\begin{equation}
\begin{aligned}
\label{eq:bathcorr}
&C_{1,\alpha}(\tau)=\sum_k\gamma_{k,\alpha}^2n_B^\alpha(\omega_{k,\alpha})e^{i(\omega_{k,\alpha}-\Omega_\alpha)\tau},\\&
C_{2,\alpha}(\tau)=\sum_k\gamma_{k,\alpha}^2\left[1+n_B^\alpha(\omega_{k,\alpha})\right]e^{-i(\omega_{k,\alpha}-\Omega_\alpha)\tau},
\end{aligned}
\end{equation}
where we assumed the baths to be in thermal states. The Bose-Einstein distribution is given by
\begin{equation}
\label{eq:boseeinstein}
n_B^\alpha(\omega)=\frac{1}{e^{\omega/(k_BT_\alpha)}-1}.
\end{equation}
The bath correlation functions are usually peaked around $\tau=0$ and decay for large times. This decay defines the bath-correlation time $\tau_B$ such that $C_{j,\alpha}(\tau_B)\ll C_{2,\alpha}(0)$ (note that $C_{1,\alpha}(0)$ vanishes as $T_\alpha\rightarrow0$). For an explicit evaluation of the bath correlation functions and a discussion on the relevance of $\tau_B$ (including a discussion of the zero temperature limit), we refer the reader to Ref.~\cite{rivas:2010}. 

The Born-Markov equation represents the starting point of our analysis. Since it does not guarantee positive evolution \cite{breuer:book}, further approximations are usually made to obtain a master equation in Gorini-Kossakowski-Sudarshan-Lindblad (GKSL) form \cite{gorini:1976,lindblad:1976} ensuring completely positive dynamics. In the following, we will discuss two popular approximations, the local and the global approach, in some detail.

\subsection{Local master equation}
The Markov approximation that was made to obtain Eq.~\eqref{eq:bornmarkovint} is responsible for the fact that the density matrix under the integral is independent of $\tau$. This approximation is valid as long as the characteristic time over which $\tilde{\rho}(t)$ varies is much larger than $\tau_B$. In the same spirit, we can make the approximation
\begin{equation}
\label{eq:localapprox}
\tilde{a}_\alpha(t-\tau)\simeq\tilde{a}_\alpha(t),
\end{equation}
in the integral of Eq.~\eqref{eq:bornmarkovint}. Note that we make this approximation in the rotating frame, where the fast oscillations with frequency $\Omega_\alpha$ are encoded in the bath correlation functions [cf.~Eq.~\eqref{eq:bathcorr}]. In our model, we have
\begin{equation}
\label{eq:ainterrot}
\tilde{a}_\alpha(t)=\hat{a}_\alpha\cos(gt)-i\hat{a}_{\bar{\alpha}}\sin(gt),
\end{equation}
where $\bar{\alpha}\neq\alpha$. The approximation in Eq.~\eqref{eq:localapprox} is therefore expected to be good as long as $g\tau_B\ll1$. For reasonably small values of $g$, this approximation is therefore completely consistent with the Markov approximation.

This approximation directly results in the local master equation (in the Schr\"odinger picture) without the need of a secular approximation
\begin{equation}
\label{eq:localmaster}
\partial_t\hat{\rho}(t)=-i[\bar{H}_S,\hat{\rho}(t)]+\sum_{\alpha=h,c}\left\{\bar{\Gamma}_\alpha\mathcal{D}[\hat{a}_\alpha]\hat{\rho}(t)+{\Gamma}_\alpha\mathcal{D}[\hat{a}^\dagger_\alpha]\hat{\rho}(t)\right\},
\end{equation}
where we neglected a physically irrelevant constant and introduced
\begin{equation}
\label{eq:gammaloc}
\bar{\Gamma}_\alpha=\kappa_\alpha(\Omega_\alpha)[n_B^\alpha(\Omega_\alpha) +1],\hspace{1cm}{\Gamma}_\alpha=\kappa_\alpha(\Omega_\alpha)n_B^\alpha(\Omega_\alpha).
\end{equation}
as well as the Lindblad superoperators
\begin{equation}
\label{eq:lindbladsuper}
\mathcal{D}[\hat{A}]\hat{\rho}=\hat{A}\hat{\rho}\hat{A}^\dagger-\frac{1}{2}\left\{\hat{A}^\dagger\hat{A},\hat{\rho}\right\},
\end{equation}
and the energy damping rate
\begin{equation}
\label{eq:spectraldens}
\kappa_\alpha(\omega)=2\pi\sum_k\gamma_{k,\alpha}^2\delta(\omega-\omega_{k,\alpha})=2\pi\rho_\alpha(\omega),
\end{equation}
where $\rho_\alpha(\omega)$ denotes the spectral density.
The renormalized Hamiltonian is given by $\bar{H}_S=\hat{H}_S+\sum_{\alpha=h,c}\Sigma_\alpha\hat{a}_\alpha^\dagger\hat{a}_\alpha$ with
\begin{equation}
\label{eq:lambshift}
\Sigma_\alpha=P\int_{0}^{\infty}d\omega\frac{\rho_\alpha(\omega)}{\Omega_\alpha-\omega},
\end{equation}
where $P$ denotes the Cauchy principal value. We note that this renormalization should be small in order for the Born-Markov approximations to be valid (this can be understood by noting that $\tilde{\rho}(t)$ will have terms that oscillate with frequency $\Sigma_\alpha$). In the following, we will thus neglect this renormalization. As shown in Sec.~\ref{sec:results}, excellent agreement between the local master equation and exact numerics is found for a wide range of parameters without taking into account the renormalization of the Hamiltonian; see also Ref.~\cite{rivas:2010}.

As discussed above, the local master equation is justified as long as $g\tau_B\ll1$. With the help of Eqs.~\eqref{eq:bathcorr}, this temporal inequality can be translated into an inequality involving the Fourier transform of the bath correlation functions. If $\tau_Bg\ll1$, then we have $\exp(ig\tau)C_{j,\alpha}(\tau)\simeq C_{j,\alpha}(\tau)$ for all $\tau$ since we can approximate $C_{j,\alpha}(\tau>\tau_B)\simeq0$. It is straightforward to show that this is fulfilled as long as the Fourier transform of the bath correlation functions can be approximated as constant over the energy scale of $g$. With the help of Eqs.~\eqref{eq:bathcorr} and \eqref{eq:spectraldens}, this results in the inequalities
\begin{equation}
\label{eq:justiflocal}
\begin{aligned}
&|n_B^\alpha(\Omega_\alpha\pm g)-n_B^\alpha(\Omega_\alpha)|\ll 1\,\,{\rm or}\,\,n_B^\alpha(\Omega_\alpha),\\
&|\kappa_\alpha(\Omega_\alpha\pm g)-\kappa_\alpha(\Omega_\alpha)|\ll\kappa_\alpha(\Omega_\alpha).
\end{aligned}
\end{equation}
Here we used the fact that the main contribution for the bath correlation functions comes from the low-frequency terms. For a bosonic bath with Ohmic spectrum, the last equations are fulfilled as long as $g\ll\Omega_\alpha$, which is usually the case for systems where the local approach is employed \cite{linden:2010prl,brunner:2012,brask:2015njp,hofer:2016prb}. For fermionic baths, equations analogous to Eqs.~\eqref{eq:justiflocal} can be derived. These are usually not fulfilled at low temperatures due to the step-like behavior of the Fermi-Dirac distribution \cite{Purkayastha2016,santos:2016}. Note that the validity of the Markov approximation, which underlies both the local as well as the global master equation, requires conditions obtained from Eq.~\eqref{eq:justiflocal} by exchanging $g$ with $\kappa_\alpha(\Omega_\alpha)$.

It is sometimes stated that the local approach is only valid for interaction strengths much smaller than the induced broadening $g\ll \kappa_\alpha(\Omega_\alpha)$ (see for instance Ref.~\cite{Purkayastha2016}). As we show in Sec.~\ref{sec:results}, the local approach gives reliable predictions even for interactions that are several times the broadening. This is in complete agreement with Eq.~\eqref{eq:justiflocal}.

We note that the local master equation is also obtained in the so-called \textit{singular coupling limit} \cite{gorini:1978,spohn:1980,breuer:book,alicki:2006}, where the bath correlation functions in Eqs.~\eqref{eq:bathcorr} tend to a delta function. This limit is often dismissed as being unrealistic. However we stress that the bath correlation functions only have to behave like delta functions on time-scales of the order of $1/g$ for the local master equation to be valid. In the weak coupling limit, which is often the regime of interest for thermal machines, $1/g$ can naturally be much bigger than $\tau_B$.

In order to solve the master equation, we make use of its bi-linearity (in creation and annihilation operators) which implies that a Gaussian state remains Gaussian at all times. Furthermore, there are no terms in the master equation which result in a displacement of the state. We can therefore restrict the analysis to states which have $\langle\hat{a}_\alpha\rangle=0$. Then the state is fully described by its covariance matrix. 
From the local master equation in Eq.~\eqref{eq:localmaster}, one can derive the following differential equations for the covariance matrix elements
\begin{equation}
\label{eq:localcovdiff}
\begin{aligned}
&\partial_t\langle \hat{a}^\dagger_h \hat{a}_h\rangle =2g{\rm Im}\left\{\langle \hat{a}^\dagger_h\hat{a}_c\rangle\right\}+\kappa_h\left(n_B^h-\langle \hat{a}^\dagger_h \hat{a}_h\rangle\right),\\
&\partial_t\langle \hat{a}^\dagger_c \hat{a}_c\rangle =-2g{\rm Im}\left\{\langle \hat{a}^\dagger_h\hat{a}_c\rangle\right\}+\kappa_c\left(n_B^c-\langle \hat{a}^\dagger_c \hat{a}_c\rangle\right),\\
&\partial_t\langle \hat{a}^\dagger_h\hat{a}_c\rangle=-\frac{(\kappa_c+\kappa_h)}{2}\langle \hat{a}^\dagger_h\hat{a}_c\rangle-ig\left(\langle \hat{a}^\dagger_h \hat{a}_h\rangle-\langle \hat{a}^\dagger_c \hat{a}_c\rangle\right),
\end{aligned}
\end{equation}
where 
\begin{equation}
\label{eq:kappanloc}
\kappa_{\alpha}=\kappa_\alpha(\Omega_{\alpha}),\hspace{1cm}n_B^{\alpha}=n_B^\alpha(\Omega_{\alpha}).
\end{equation}
In the steady state, these differential equations are solved by the time-independent expressions
\begin{equation}
\label{eq:covsteadyloc}
\begin{aligned}
&\langle \hat{a}^\dagger_h \hat{a}_h\rangle =n_B^h-\frac{4g^2\kappa_c(n_B^h-n_B^c)}{(\kappa_h+\kappa_c)(\kappa_c\kappa_h+4g^2)},\\
&\langle \hat{a}^\dagger_c \hat{a}_c\rangle =n_B^c+\frac{4g^2\kappa_h(n_B^h-n_B^c)}{(\kappa_h+\kappa_c)(\kappa_c\kappa_h+4g^2)},\\
&\langle \hat{a}^\dagger_h\hat{a}_c\rangle=\frac{-i2g\kappa_c\kappa_h(n_B^h-n_B^c)}{(\kappa_h+\kappa_c)(\kappa_c\kappa_h+4g^2)}.
\end{aligned}
\end{equation}

\subsection{Global master equation}

The second approximation we consider is the secular approximation. This approximation consists of dropping all the terms in the Born-Markov master equation [cf.~Eq.~\eqref{eq:bornmarkovint}] which oscillate as a function of time $t$. The secular approximation is expected to hold over time-scales much bigger than the inverse frequencies of the oscillating terms and obviously becomes better as these frequencies increase. In order to identify the oscillating terms, we write the interaction picture operators in Eq.~\eqref{eq:ainterrot} as
\begin{equation}
\label{eq:aint}
\tilde{a}_h(t)=\frac{1}{\sqrt{2}}\left(\hat{a}_+e^{-igt}+\hat{a}_-e^{igt}\right),\hspace{.75cm}
\tilde{a}_c(t)=\frac{1}{\sqrt{2}}\left(\hat{a}_+e^{-igt}-\hat{a}_-e^{igt}\right),
\end{equation}
where we introduced the operators
\begin{equation}
\label{eq:opsigma}
\hat{a}_\pm=\frac{1}{\sqrt{2}}\left(\hat{a}_h\pm\hat{a}_c\right).
\end{equation}

The secular approximation is obtained by plugging Eq.~\eqref{eq:aint} into Eq.~\eqref{eq:bornmarkovint} and dropping all terms that oscillate with $\exp[2igt]$. 
This results in the master equation
\begin{equation}
\label{eq:mastersecdegen}
\partial_t\hat{\rho}(t)=-i[\bar{H}_S,\hat{\rho}(t)]
+\frac{1}{2}\sum_{\sigma=\pm}\left\{\bar{\Gamma}_\sigma\mathcal{D}[\hat{a}_\sigma]\hat{\rho}(t)+\Gamma_\sigma\mathcal{D}[\hat{a}_\sigma^\dagger]\hat{\rho}(t)\right\},
\end{equation}
where
\begin{equation}
\label{eq:gammasecd}
\bar{\Gamma}_\sigma=\sum_{\alpha=h,c}\kappa_{\alpha}(\Omega_{\alpha,\sigma})[n_B^{\alpha}(\Omega_{\alpha,\sigma}) +1],\hspace{.75cm}\Gamma_\sigma=\sum_{\alpha=h,c}\kappa_{\alpha}(\Omega_{\alpha,\sigma}) n_B^{\alpha}(\Omega_{\alpha,\sigma}),
\end{equation}
with the frequencies
\begin{equation}
\label{eq:omalsig}
\Omega_{\alpha,\pm}=\Omega_\alpha\pm g.
\end{equation}
The renormalized Hamiltonian reads
\begin{equation}
\label{eq:hamrenormsec}
\bar{H}_S=\sum_{\sigma=\pm}\left(\sigma g+\Sigma_{h,\sigma}+\Sigma_{c,\sigma}\right)\hat{a}_\sigma^\dagger\hat{a}_\sigma,
\end{equation}
and
\begin{equation}
\label{eq:lambshiftdeg}
\Sigma_{\alpha,\sigma}=\frac{1}{2}P\int_{0}^{\infty}d\omega\frac{\rho_\alpha(\omega)}{\Omega_{\alpha,\sigma}-\omega}.
\end{equation}
As for the local approach, we will neglect the renormalization of the Hamiltonian in the following.

The secular approximation results in a master equation where the eigenmodes of the Hamiltonian $\hat{a}_\sigma$ are the relevant degrees of freedom. The action of the baths decouples into a bath for each eigenmode, which drives the mode towards thermal equilibrium characterized by the occupation numbers
\begin{equation}
\label{eq:eigenocc}
n_\sigma = \frac{\kappa_{h,\sigma}n_B^{h,\sigma}+\kappa_{c,\sigma}n_B^{c,\sigma}}{\kappa_{h,\sigma}+\kappa_{c,\sigma}},
\end{equation}
where we introduced
\begin{equation}
\label{eq:kappanglob}
\kappa_{\alpha,\sigma}=\kappa_\alpha(\Omega_{\alpha,\sigma}),\hspace{1cm}n_B^{\alpha,\sigma}=n_B^\alpha(\Omega_{\alpha,\sigma}).
\end{equation}
We note that in the limit $g\rightarrow 0$, the local master equation is no longer recovered. Because the secular approximation is no longer justified in this limit, we expect the global master equation to break down. This is also consistent with the fact that we expect the secular approximation to be valid as long as the frequency of the neglected oscillating terms is much bigger than the linewidths, i.e. $\kappa_\alpha\ll g$. Since the Markov approximation requires $\kappa_\alpha\ll\Omega_\alpha$, and the local approach is valid for $g\ll\Omega_\alpha$, there is an overlap between the regimes of validity of the local and the global master equation. This implies that the local and the global master equation together are enough to describe the system for all parameters which allow for a Markovian master equation.

In the global approach, the covariance matrix is governed by the differential equations
\begin{equation}
\label{eq:covglob}
\begin{aligned}
&\partial_t\langle \hat{a}^\dagger_\sigma \hat{a}_\sigma\rangle = \frac{1}{2}\sum_{\alpha=h,c}\kappa_{\alpha,\sigma}\left[n_B^{\alpha,\sigma}-\langle \hat{a}^\dagger_\sigma \hat{a}_\sigma\rangle\right],\\&
\partial_t\langle \hat{a}^\dagger_+ \hat{a}_-\rangle =\left[2ig-\frac{1}{4}\sum_{\alpha,\sigma}\kappa_{\alpha,\sigma}\right]\langle \hat{a}^\dagger_+ \hat{a}_-\rangle.
\end{aligned}
\end{equation}
In the steady state, we find that the state is a product of thermal states with occupation numbers $\langle \hat{a}^\dagger_\pm \hat{a}_\pm\rangle=n_\sigma$ as expected.

\section{Exact numerics}
\label{sec:numerics}
In this section we briefly describe how we obtain exact numerics which are used as a benchmark when comparing the different master equations. To this end, we simulate the unitary evolution generated by the Hamiltonian in Eq.~\eqref{eq:hamiltonian} for big but finite baths. The key element here is that the Hamiltonian in Eq.~\eqref{eq:hamiltonian} is quadratic, so that Gaussian states (such as thermal states) remain Gaussian throughout the whole evolution. As such, they can be fully characterized by only their first and second moments (see, e.g., \cite{eisert2003}). This allows us to characterize the time-evolved state of the whole system, including the thermal baths, by a matrix of size $\sim N$, where $N$ is the total number of oscillators involved. 

We consider baths made up of $n+1$ harmonic oscillators, so that the total size of system and baths is $N=2(n+2)$. The bath modes are chosen to be uniformly spread over a range $(0,\omega_{\rm c})$, where $\omega_{\rm c}$ is a cutoff frequency. That is,
\begin{align}
\label{eq:DiscreteFrequencies}
\omega_{k,\alpha}=\frac{k}{n}\omega_{\rm c},
\end{align}
for $k=0,..,n$. This defines $\hat{H}_{\alpha}$ in Eq.~\eqref{eq:hamiltonian}. Let us now turn our attention to $\hat{V}_{\alpha}$, and hence to the couplings $\gamma_{k,\alpha}$. First note that the action of the baths in the Markovian master equations, Eqs.~\eqref{eq:lindbladsuper} and \eqref{eq:mastersecdegen}, is captured by the spectral density given in Eq.~\eqref{eq:spectraldens}. It is then common to use an \emph{ad hoc} form for the spectral density in the continuum limit, instead of specifying the coupling constants $\gamma_{k,\alpha}$. A common choice for the spectral density is an Ohmic spectrum,
\begin{equation}
\rho_{\alpha}(\omega) \propto \omega,
\end{equation}
which holds for low frequencies, $\omega\leq \omega_{\rm c}$. In order to relate this approach to the $\gamma_{k,\alpha}$'s, which are necessary to simulate the full Hamiltonian in the finite-baths scenario, let us integrate Eq.~\eqref{eq:spectraldens}, obtaining,
\begin{align}
\int_{0}^{\omega_{\rm c}} \rho_{\alpha} (\omega) d\omega= \sum_{k=0}^n \gamma^2_{k,\alpha}.
\end{align}
It is now convenient to discretize the integral,
\begin{align}
\int_{0}^{\omega_{\rm c}} \rho_{\alpha} (\omega) d\omega \approx \sum_{k=0}^n \rho_{\alpha}(\omega_{k,\alpha}) \frac{\omega_c}{n},
\end{align}
from where it immediately follows,
\begin{align}
\label{eq:DiscreteInteractions}
\gamma^2_{k,\alpha} \approx \rho_{\alpha}(\omega_{k,\alpha}) \frac{\omega_c}{n},
\end{align}
which becomes increasingly accurate with increasing $n$. In the particular case of an Ohmic distribution, we obtain,
\begin{align}
\gamma_{k,\alpha} \propto \sqrt{k} \frac{\omega_c}{n}.
\end{align}
Equation \eqref{eq:DiscreteInteractions}, together with Eq.~\eqref{eq:DiscreteFrequencies}, provides a simple recipe for building the discrete version of the Hamiltonian in Eq.~\eqref{eq:hamiltonian} for a given spectral density. 
The specific choice of $\omega_c$ and $n$ for our simulations, as well as the dependence of the results on this choice, is discussed in \ref{app:numerics}.

The initial state for the simulations is taken to be of the form,
\begin{align}
\hat{\rho}_0= \hat{\tau}_{\beta_h} \otimes \hat{\rho}_S \otimes \hat{\tau}_{\beta_c}
\label{initialstatesimulations}
\end{align}
where $\hat{\tau}_{\beta_{\alpha}}$ are thermal states,
\begin{align}
\hat{\tau}_{\beta_{\alpha}}=\frac{e^{-\beta_\alpha\hat{H}_{\alpha}^{l}}}{\mathcal{Z}_{B,\alpha}},
\end{align}
with inverse temperature $\beta_\alpha=1/(k_BT_\alpha)$ and we note that $\hat{H}_{\alpha}^{l}$ is in the laboratory frame. Here, the initial state of the machine $\hat{\rho}_S$ can be an arbitrary Gaussian state. In our simulations, we take $\hat{\rho}_S=e^{-\beta_h\Omega_h \hat{a}^\dagger_h\hat{a}_h}/\mathcal{Z}_h \otimes e^{-\beta_c\Omega_c \hat{a}^\dagger_c\hat{a}_c}/\mathcal{Z}_c$. 
We note that with this choice, the state given in Eq.~\eqref{initialstatesimulations} is Gaussian. 

Once the Hamiltonian and the initial state are defined, we consider the closed unitary dynamics of the full compound under the Hamiltonian in Eq.~\eqref{eq:hamiltonian} (it is convenient to work in the rotating frame, where the Hamiltonian is time independent).
The dynamics can be derived by considering the Heisenberg equations of motion of the operators $\hat{a}_{\alpha}, \hat{a}_{\alpha}^{\dagger}, \{\hat{b}_{k,\alpha}, \hat{b}^{\dagger}_{k,\alpha} \}$. For details on the derivation, we refer the reader to Ref.~\cite{rivas:2010}.

In order to simulate the steady state using finite degrees of freedom, one needs to let the whole compound evolve for a time $t$ that satisfies $\tau_{\rm eq} \ll t \ll \tau_{\rm rec}$, where $\tau_{\rm eq}$ is the equilibration time of the system and $\tau_{\rm rec}$ is the recurrence time of the bath. From Eqs.~\eqref{eq:localmaster} and \eqref{eq:localcovdiff}, we infer the equilibration time to be $1/\tau_{\rm eq} \approx \max\{\kappa_h , \kappa_c\}$ (see also \cite{perarnau:2017}, where the equilibration time is discussed explicitly for finite systems). On the other hand, the recurrence time scales linearly with the number of oscillators in the bath \cite{rivas:2010}. Hence, by taking a sufficiently large bath (in the simulations we take $\sim 400$ oscillators), we can ensure that $\tau_{\rm eq} \ll \tau_{\rm rec}$. In our simulations, we take $t\approx 20 \tau_{\rm eq}$.

\section{Observables and reduced states}
\label{sec:obs}
In this work, we are particularly interested in the energy flows that traverse the quantum thermal machine in a non-equilibrium situation. However, to compare to previous works \cite{rivas:2010}, and to further assess the validity of the different master equations, we also consider the obtained steady states.

\subsection{Heat currents, power, and efficiency}

\subsubsection{Local master equation}
As we consider a heat engine, the main quantity of interest is the power that is produced. For our system, it is defined as \cite{vinjanampathy:2016}
\begin{equation}
\label{eq:power}
P=-{\rm Tr}\left\{ [\partial_t \hat{H}^l_S(t)]\hat{\rho}^l(t)\right\} = -2g\mathcal{E}{\rm Im}\left\{\left\langle\hat{a}_h^\dagger\hat{a}_c\right\rangle\right\},
\end{equation}
where the superscript denotes the laboratory frame and $\langle\cdots\rangle$, denotes the ensemble average in the rotating frame.
Note that positive power implies that energy leaves the system.
In addition to the power, we consider the heat currents that enter (or leave) the system. To this end, we write
\begin{equation}
\label{eq:mastershort}
\partial_t\hat{\rho}^l=-i[\hat{H}^l_S(t),\hat{\rho}^l(t)]
+\sum_{\alpha=h,c}\mathcal{L}^l_\alpha\hat{\rho}^l(t),
\end{equation}
where $\mathcal{L}^l_\alpha$ is a super-operator that groups all the dissipative terms which arise from bath $\alpha$ in the laboratory frame. Note that under the Born-Markov approximation, the dissipators of different baths can be added \cite{kolodynski:2017}.
The heat currents are then defined as
\begin{equation}
\label{eq:heatcurrmaster}
J_\alpha={\rm Tr}\left\{\hat{H}^l_S(t)\mathcal{L}^l_\alpha\hat{\rho}^l\right\}={\rm Tr}\left\{\bigg[\hat{H}_S+\sum_{\alpha=h,c}\Omega_\alpha \hat{a}_\alpha^\dagger\hat{a}\bigg]\mathcal{L}_\alpha\hat{\rho}\right\},
\end{equation}
 where the dissipator in the rotating frame is related to the dissipator in the laboratory frame by
\begin{equation}
\label{eq:dissprot}
\mathcal{L}^l_\alpha\hat{\rho}^l(t)=\hat{U}^\dagger_r(t)\left[\mathcal{L}_\alpha\hat{\rho}(t)\right]\hat{U}_r(t).
\end{equation}
With our sign convention, a positive heat current implies energy entering the system from the bath. In the steady state, the first law of thermodynamics thus reads
\begin{equation}
\label{eq:firstlaw}
P=J_c+J_h.
\end{equation}

The efficiency is defined as the ratio of the obtained power, divided by the heat that originates from the hot bath
\begin{equation}
\label{eq:efficiency}
\eta=\frac{P}{J_h}.
\end{equation}
In the regime where the system operates as a heat engine (i.e. $P>0$ and $J_h>0$), the second law of thermodynamics forces the efficiency to remain below the Carnot limit
\begin{equation}
\label{eq:secondlaw}
\eta<1-\frac{T_c}{T_h}=\eta_C.
\end{equation}

For the local master equation in Eq.~\eqref{eq:localmaster}, the heat currents in Eq.~\eqref{eq:heatcurrmaster} can be written as
\begin{equation}
\label{eq:heatcurrlocal}
J_\alpha=\kappa_\alpha\Big[{\Omega}_\alpha\left(n_B^\alpha-\langle \hat{a}^\dagger_\alpha \hat{a}_\alpha\rangle\right)-\frac{g}{2}\langle \hat{a}^\dagger_h\hat{a}_c+\hat{a}^\dagger_c\hat{a}_h\rangle\Big].
\end{equation}
From Eqs.~\eqref{eq:localcovdiff}, we can infer that the second term in the heat current decays exponentially in time. In the steady state, from Eqs.~\eqref{eq:covsteadyloc} and \eqref{eq:power}, we find for the power
\begin{equation}
\label{eq:powerloc}
P=\frac{(\Omega_h-\Omega_c)4g^2\kappa_c\kappa_h(n_B^h-n_B^c)}{(\kappa_h+\kappa_c)(\kappa_c\kappa_h+4g^2)},
\end{equation}
and the heat current
\begin{equation}
\label{eq:heatcurrloc}
J_h={\Omega}_h\frac{4g^2\kappa_c\kappa_h(n_B^h-n_B^c)}{(\kappa_h+\kappa_c)(\kappa_c\kappa_h+4g^2)},
\end{equation}
resulting in the efficiency
\begin{equation}
\label{eq:effloc}
\eta=1-\frac{\Omega_c}{\Omega_h}.
\end{equation}
We note that this efficiency fulfills Eq.~\eqref{eq:secondlaw}. When the frequencies are chosen such that the efficiency is above the Carnot efficiency, then we find $P<0$ and $J_h<0$. As long as $\eta<\eta_C$, our machine is thus a heat engine, $\eta=\eta_C$ denotes the point of reversibility, where all the energy currents vanish, and for $\eta>\eta_C$, the machine acts as a refrigerator, using power to induce a heat current from the cold bath to the hot bath \cite{hofer:2016prb}. Finally, we note that for $\Omega_h=\Omega_c$, where there is no external power, heat always flows from the hot bath to the cold bath as dictated by the second law of thermodynamics. In contrast to models which consider non-energy preserving interactions, the local approach does not violate the laws of thermodynamics when including an external field that provides the energy to convert photons of frequency $\Omega_c$ into photons of frequency $\Omega_h$.

\subsubsection{Global master equation}\label{sec:globalheat}
In the global master equation, the bath couples to global states which are dressed by the external field. Therefore, the dissipative terms include the external field and the definitions introduced in the last subsection are no longer valid. To define heat and work in the global approach, we follow Refs.~\cite{alicki:2006,levy:2012pre,szczygielski:2013,kosloff:2013,gelbwaser:2015}. To this end, we first write the dissipator in the global master equation [cf.~Eq.~\eqref{eq:mastersecdegen}] as the sum of four dissipators given by
\begin{equation}
\label{eq:dissipatorsum}
\mathcal{L}_{\alpha,\sigma}\hat{\rho}(t)=\frac{1}{2}\kappa_{\alpha,\sigma}(n_B^{\alpha,\sigma}+1)\mathcal{D}[\hat{a}_\sigma]\hat{\rho}(t)+
\frac{1}{2}\kappa_{\alpha,\sigma}n_B^{\alpha,\sigma}\mathcal{D}[\hat{a}^\dagger_\sigma]\hat{\rho}(t).
\end{equation}
We further introduce the thermal states
\begin{equation}
\label{eq:statesdiss}
\hat{\rho}_{\alpha,\sigma}=\frac{e^{-\beta_\alpha \Omega_{\alpha,\sigma}\hat{a}_\sigma^\dagger\hat{a}_\sigma}}{\mathcal{Z}_{\alpha,\sigma}},
\end{equation}
which constitute the steady states of the respective operators, i.e. $\mathcal{L}_{\alpha,\sigma}\hat{\rho}_{\alpha,\sigma}=0$.

The heat currents in the steady state (denoted $\hat{\rho}$) are then defined as 
\begin{equation}
\label{eq:heatcurrglob}
J_\alpha=-k_BT_\alpha\sum_{\sigma=\pm}{\rm Tr}\left\{\left[\mathcal{L}_{\alpha,\sigma}\hat{\rho}\right]\ln\hat{\rho}_{\alpha,\sigma}\right\},
\end{equation}
and the power is given by the first law, cf.~Eq.~\eqref{eq:firstlaw}. We note that for the global approach, we enforce the first law while in the local approach, it follows from the expressions for heat currents and power.

An explicit calculation results in \cite{levy:2012pre}
\begin{equation}
\label{eq:heatcurrglobexpl}
J_h=\frac{1}{2}\sum_{\sigma=\pm}\Omega_{h,\sigma}\frac{\kappa_{h,\sigma}\kappa_{c,\sigma}}{\kappa_{h,\sigma}+\kappa_{c,\sigma}}(n_B^{h,\sigma}-n_B^{c,\sigma}).
\end{equation}
Note that if Eqs.~\eqref{eq:justiflocal} are fulfilled, the last expression reduces to
\begin{equation}
\label{eq:heatcurrglobloc}
J_h=\Omega_h\frac{\kappa_h\kappa_c}{\kappa_h+\kappa_c}(n_B^{h}-n_B^{c}),
\end{equation}
which is the expression obtained by the local approach in the limit $g\gg\kappa_\alpha$. As expected, if the approximations leading to both the local and the global master equation are justified, the two approaches give the same result. Whenever Eqs.~\eqref{eq:justiflocal} are not fulfilled, the global approach implies that the heat engine makes use of two channels labeled by the subscript $\sigma$. Each of these channels has an efficiency
\begin{equation}
\label{eq:etachannel}
\eta_\sigma=\frac{P_\sigma}{J_{h,\sigma}}=\frac{J_{h,\sigma}+J_{c,\sigma}}{J_{h,\sigma}}=1-\frac{\Omega_{c,\sigma}}{\Omega_{h,\sigma}}\leq\eta_C,
\end{equation}
where $J_\alpha=J_{\alpha,+}+J_{\alpha,-}$ and $J_c$ is obtained from $J_h$ by exchanging the labels $c\leftrightarrow h$ [cf.~Eq.~\eqref{eq:heatcurrglobexpl}]. We focus on the heat engine regime where $J_{h,\sigma}\,,P_{\sigma}\,>0$. The total efficiency is then of the form $\eta=c\eta_++(1-c)\eta_-$, with $c=J_{h,+}/(J_{h,+}+J_{h,-})\leq 1$. The efficiency can only reach the Carnot value if one of the channels carries no energy (i.e. $c=1$ or $c=0$) or if both channels can simultaneously reach the Carnot point (which is the case if Eqs.~\eqref{eq:justiflocal} hold). The fact that heat engines can only reach Carnot efficiency if they are effectively reduced to a single channel is also discussed in Ref.~\cite{brunner:2012}.

\subsubsection{Exact numerics}

When dealing with the full Hamiltonian, we define heat currents as the energy lost by the bath,
\begin{align}
J_{\alpha} = - \frac{d {\rm Tr}\{\hat{H}_{\alpha}^{l} \hat{{\varrho}}^l(t)\}}{dt},
\end{align}
where we note that both $\hat{H}_{\alpha}^{l}$ and $\hat{\varrho}^l(t)$, the unitarily time evolved state of the whole compound, are taken in the lab frame. The power can be obtained through Eq.~\eqref{eq:power}.  

\subsection{Reduced states}

In addition to the energy currents, we consider the reduced state of the system as obtained by the different solutions. Since the considered Hamiltonian is bi-linear in bosonic annihilation and creation operators, it suffices to consider the covariance matrices. It will be convenient to work in the $\hat{x}$, $\hat{p}$ basis, with $\hat{x}_{\alpha}=\sqrt{1/2\Omega_{\alpha}}(\hat{a}^{\dagger}_{\alpha}+\hat{a}_{\alpha})$ and $\hat{p}_{\alpha}=i\sqrt{\Omega_{\alpha}/2}(\hat{a}^{\dagger}_{\alpha}-\hat{a}_{\alpha})$. Defining the vector $\hat{\bf r}=(\hat{x}_h,\hat{x}_c,\hat{p}_h,\hat{p}_c)$, the covariance matrix is given by,
\begin{align}
\mathcal{C}_{ij}=\frac{1}{2}{\rm Tr}\left(\hat{\rho}(\hat{r}_i \hat{r}_j+\hat{r}_j \hat{r}_i) \right).
\end{align}
For the master equations, the covariance matrix can be obtained straightforwardly from Sec.~\ref{sec:mastereqs} [cf.~Eqs.~\eqref{eq:covsteadyloc}, and the discussion around Eq.~\eqref{eq:covglob}]. In the case of the exact numerics, one needs to consider the relevant entries of the covariance matrix of the whole compound (see, e.g., Ref~\cite{rivas:2010}).

In the next section, we compare the reduced states obtained from the master equations to the reduced state obtained through exact numerics. As a measure of distinguishability between two states $\hat{\rho}$ and $\hat{\sigma}$, we consider the fidelity, defined as,
\begin{equation}
\mathcal{F}(\hat{\rho},\hat{\sigma})={\rm Tr}\left(\sqrt{\sqrt{\hat{\rho}} \hat{\sigma} \sqrt{\hat{\rho}}} \right).
\end{equation}
For two-mode Gaussian states, represented by covariance matrices $\mathcal{C}_{1}$ and $\mathcal{C}_{2}$, and vanishing first order moments, the fidelity is given by \cite{Marian2012},
\begin{equation}
\mathcal{F}(\mathcal{C}_{1},\mathcal{C}_{2})=\left[\sqrt{b}+\sqrt{c}-\sqrt{(\sqrt{b}+\sqrt{c})^2-a} \right]^{-1}.
\end{equation}
Here $a=\det(\mathcal{C}_{1}+\mathcal{C}_{2})$, $b=2^4 \det(J \mathcal{C}_{1} J \mathcal{C}_{2}-\mathbb{I}/4)$, $c=2^4 \det(\mathcal{C}_{1}+iJ/2) \det(\mathcal{C}_2+iJ/2)$, and the matrix elements of $J$ are given by $J_{kl}=-i\langle [\hat{r}_k, \hat{r}_l]\rangle$.

\section{Results}
\label{sec:results}

\begin{figure}[t!]
\centering
\includegraphics[width=\textwidth]{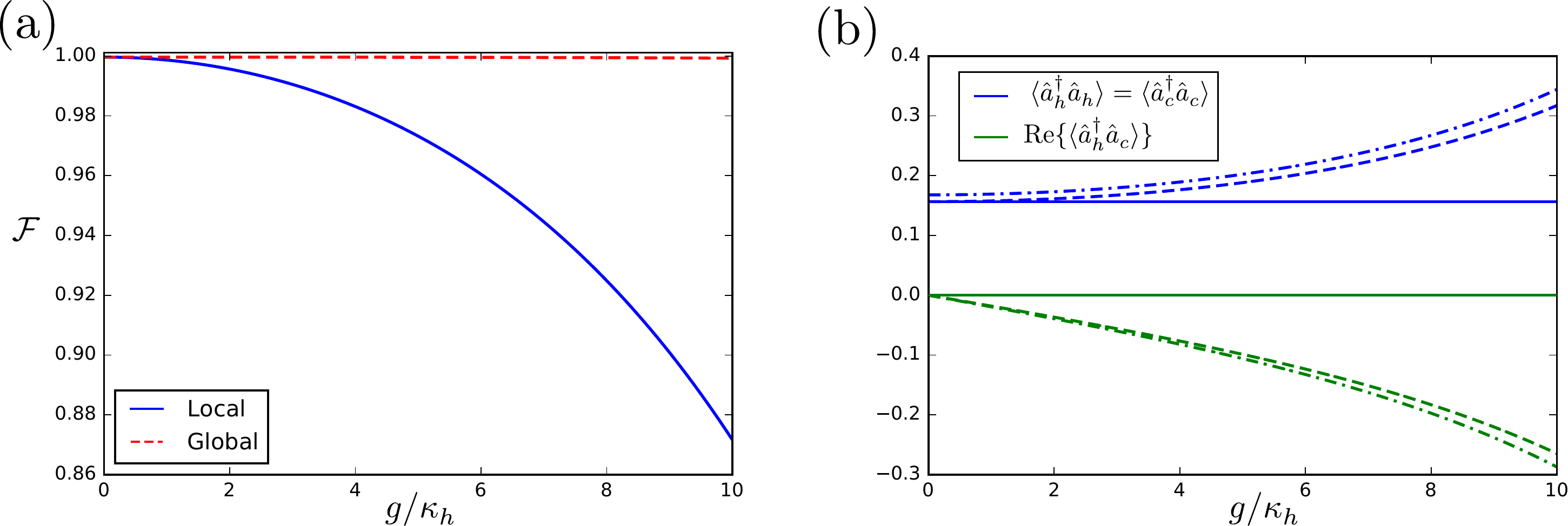}
\caption{Comparison of steady states in equilibrium obtained from the local master equation [cf.~Eq.~\eqref{eq:localmaster}], the global master equation [cf.~Eq.~\eqref{eq:mastersecdegen}], and exact numerics. ${\rm (a)}$ Fidelity between the states obtained from the master equation and the state obtained from exact numerics. ${\rm (b)}$ Other observables as a function of interaction strength. In the equilibrium case, ${\rm Im}\{\hat{a}_h^\dagger\hat{a}_c\}=0$. Solid: local master equation, dashed: global master equation, dash-dotted: exact numerics. The global master equation always outperforms the local master equation, except for the limit $g\rightarrow0$, where the two approaches result in the same steady state. 
Parameters: $\Omega_h=\Omega_c=1$, $\kappa_h=\kappa_c=0.05$, $k_BT_c=k_BT_h=0.5$. Parameters numerics: $\omega_c=3$, $n=400$, $t=20/\kappa$, where $t$ is the time we let the whole compound evolve to equilibrate.}
  \label{fig:equilibrium}
\end{figure}

\begin{figure}[t!]
\centering
\includegraphics[width=\textwidth]{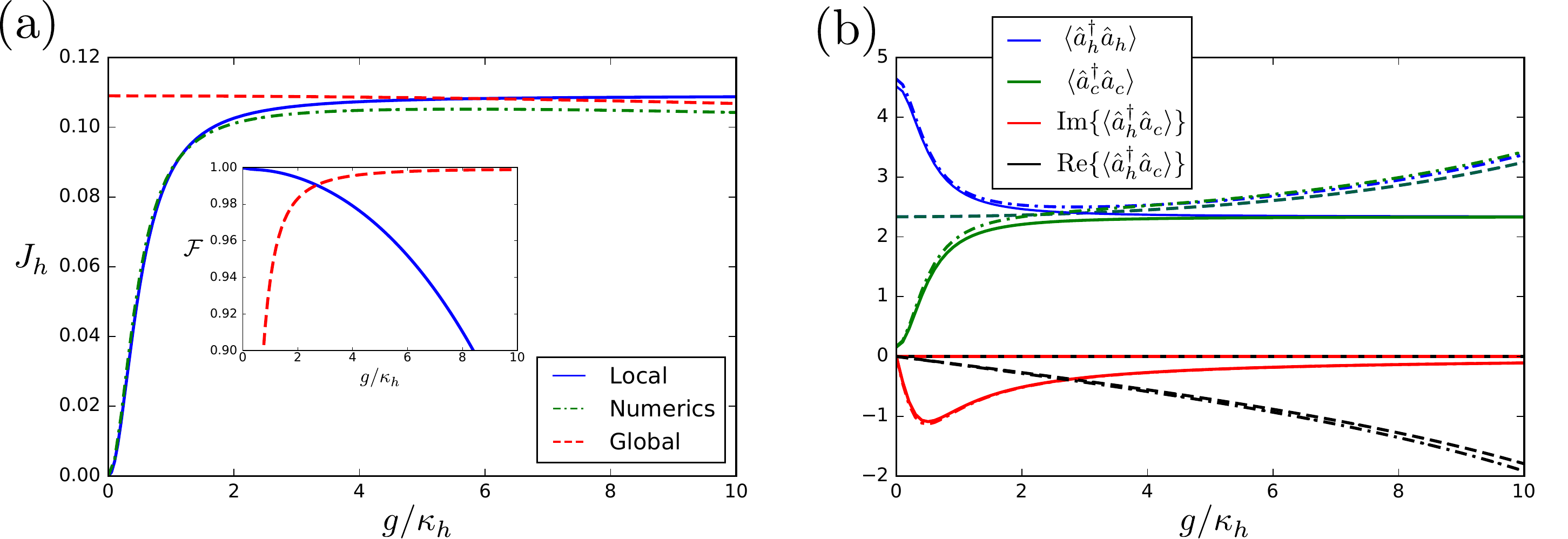}
\caption{Comparison of heat currents and other observables obtained from the local master equation [cf.~Eq.~\eqref{eq:localmaster}], the global master equation [cf.~Eq.~\eqref{eq:mastersecdegen}], and exact numerics. ${\rm (a)}$ Heat current as a function of interaction strength. The local master equation performs very well even up to interaction strength $g=\Omega_c/2=10\kappa_h$. The global master equation breaks down for small $g$ where the secular approximation is no longer justified. For higher values of $g$, the global master equation yields similar results to the local master equation and the exact numerics. The inset shows the fidelity between the state obtained from the master equations and the state obtained from exact numerics. ${\rm (b)}$ Other observables as a function of interaction strength. Solid: local master equation, dashed: global master equation, dash-dotted: exact numerics. Again we observe the breakdown of the global approach for small $g$.
Parameters: $\Omega_h=\Omega_c=1$, $\kappa_h=\kappa_c=0.05$, $k_BT_c=0.5$, $k_BT_h=5$. Parameters numerics: $\omega_c=3$, $n=400$, $t=20/\kappa$.}
  \label{fig:degen_g}
\end{figure}

Our results for the heat engine are illustrated in Figs.~\ref{fig:equilibrium}-\ref{fig:power_oh}. We divide our results into three regimes. First we discuss the equilibrium regime, where the global approach shows excellent agreement with numerics for all values of the inter-system interaction $g$. Then we discuss the presence of a thermal bias but no external field. In this case, the global approach breaks down for small values of $g$. Finally, we focus on the heat engine regime which requires both a thermal bias as well as an external field. Again, the global approach breaks down for small values of $g$ as expected. In all regimes, the local approach performs well for $g\ll\Omega_\alpha$ (for both $\alpha=c,h$), which is where Eqs.~\eqref{eq:justiflocal} are fulfilled for our system. As expected, the global approach performs well for $g\gg\kappa_\alpha$ (for both $\alpha=c,h$), which is where the secular approximation is well justified. In the following, all energies are given in units of $\Omega_c$ and all energy currents are given in units of $\Omega_c^2$.

\subsection{Equilibrium}

We first present our results for the equilibrium case, where $T_c=T_h$ and $\Omega_h=\Omega_c$. \figref{fig:equilibrium}\,${\rm (a)}$ shows the fidelities between the steady states obtained from Eqs.~\eqref{eq:localmaster} and \eqref{eq:mastersecdegen}, and the steady state obtained from exact numerics. As expected, the global approach yields an accurate description of the steady state while the local approach gets progressively worse as the inter-system interaction $g$ increases. The same conclusions can be drawn from Fig.~\ref{fig:equilibrium}\,${\rm (b)}$, where observables such as occupation numbers are plotted. We note that Fig.~\ref{fig:equilibrium} goes up to $g=\Omega_c/2$ and thus covers interaction strengths which are much higher than the ones usually considered when the local master equation is employed.

We note that in the limit $g\rightarrow0$, the local and the global master equations result in the same steady state which is given by a product of thermal states with respect to the local oscillator Hamiltonians. If we also have $\kappa_c=\kappa_h$, the two master equations coincide. For energy damping rates that differ, i.e. $\kappa_c\neq\kappa_h$, we expect the two approaches to result in different predictions for the transient regime. 

These results confirm that in an equilibrium situation, the global approach is indeed preferable over the local approach, at least when one is interested in the steady state properties. One might be tempted to believe that this conclusion carries over to the out-of-equilibrium regime. That this is not the case is illustrated below.

\subsection{Thermal bias}
We now turn to the case of a thermal bias $T_c\neq T_h$, but still no external field, i.e. $\Omega_h=\Omega_c$. Our results for this regime are illustrated in Fig.~\ref{fig:degen_g}. The heat current is plotted in Fig.~\ref{fig:degen_g}\,${\rm (a)}$. For all models, we find $J_h=-J_c$ (first law) and $J_h\geq0$ (second law). As discussed above, the global approach breaks down in this case for interactions $g\lesssim\kappa_\alpha$. In this limit, the secular approximation is no longer justified and the global approach gives the unphysical result of a finite heat current in the limit $g\rightarrow 0$. The local approach on the other hand predicts the heat current extremely well up to $g=\Omega_c/2$.

The inset of Fig.~\ref{fig:degen_g}\,${\rm (a)}$ shows the fidelities with respect to the numerical solution. As expected, the local approach reliably reproduces the steady state for small values of $g$ while the global approach works well for large values of $g$. Figure \ref{fig:degen_g}\,${\rm (b)}$ shows other observables such as the occupation numbers, leading to the same conclusions. Note that the occupation numbers obtained from the local approach differ quite a bit from the ones obtained from exact numerics for large interactions. Nevertheless, the heat current is still captured very well by the local approach.

\begin{figure}[t!]
\centering
\includegraphics[width=\textwidth]{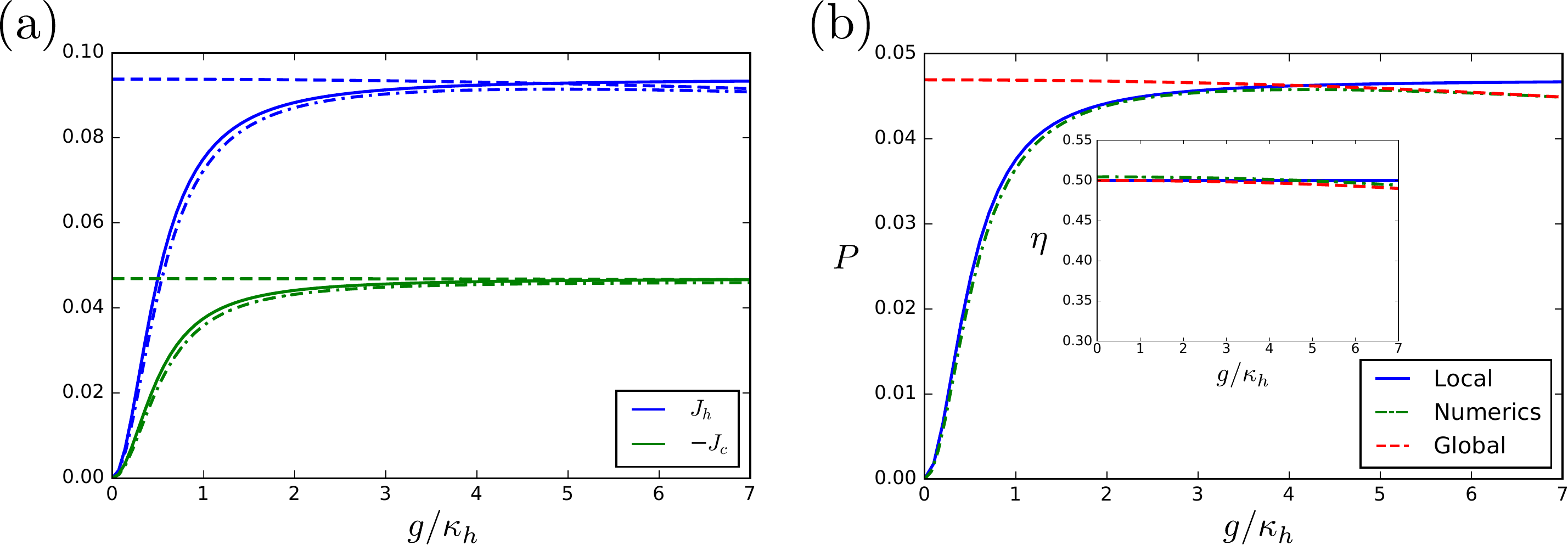}
\caption{Comparison of different models for the heat engine. ${\rm (a)}$ Heat currents: while the local approach shows very good agreement with numerics, the global approach breaks down when $g\lesssim \kappa_\alpha$. ${\rm (b)}$ Power and efficiency. Again we find excellent agreement between the local approach and exact numerics. In particular, the numerical value for the efficiency is very close to the universal value $\eta=1-\Omega_c/\Omega_h$ obtained from the local master equation. Note that when $g$ becomes comparable to $\Omega_c$, the local approach starts to deviate from the exact numerics and the global approach becomes preferable. Parameters: $\Omega_c=1$, $\Omega_h=2$ $\kappa_h=\kappa_c=0.05$, $k_BT_c=0.5$, $k_BT_h=5$. Parameters numerics: $\omega_c=3$, $n=400$, $t=20/\kappa$.}
  \label{fig:hat_engine_g}
\end{figure}

\begin{figure}[t!]
\centering
\includegraphics[width=\textwidth]{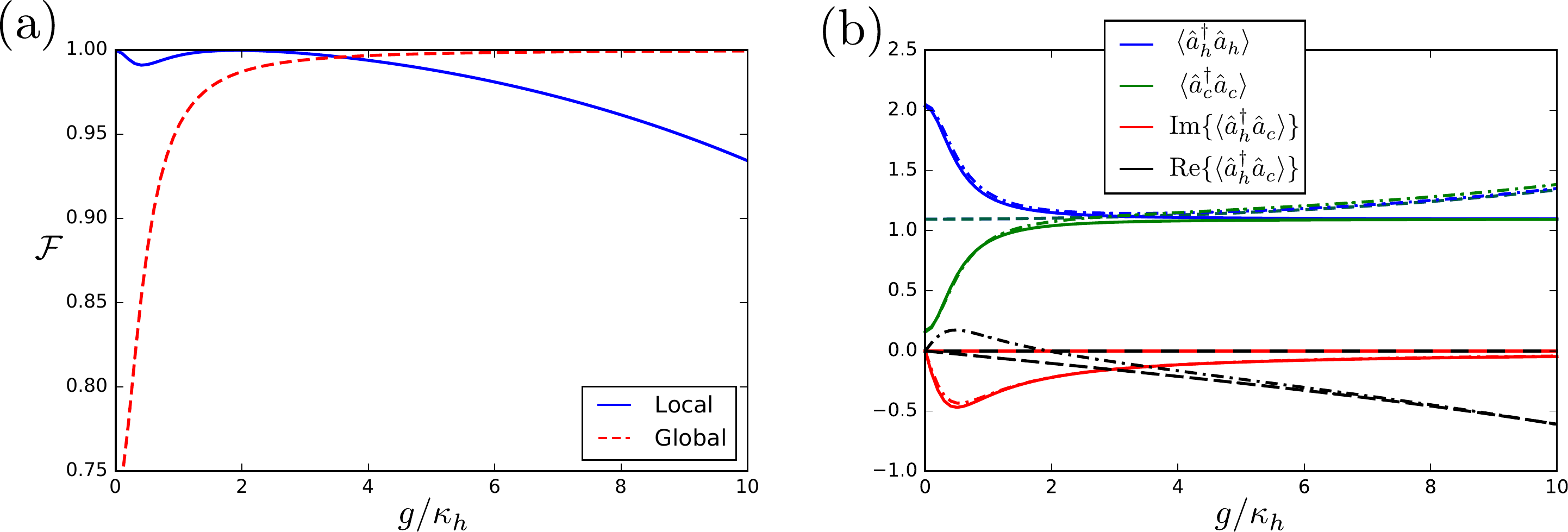}
\caption{Comparison of steady states out of equilibrium obtained from the local master equation [cf.~Eq.~\eqref{eq:localmaster}], the global master equation [cf.~Eq.~\eqref{eq:mastersecdegen}], and exact numerics. ${\rm (a)}$ Fidelity between the states obtained from the master equation and the state obtained from exact numerics. The dip in the fidelity for the local approach occurs at $g\approx|\Sigma_h|$ and is therefore assumed to arise from neglecting the renormalization of the Hamiltonian; see Eq.~\eqref{eq:lambshift}. ${\rm (b)}$ Other observables as a function of interaction strength. Solid: local master equation, dashed: global master equation, dash-dotted: exact numerics. The global approach breaks down for small values of $g$. However, for large $g$, it gives a better prediction of the steady state than the local approach.
Parameters: $\Omega_h=2$, $\Omega_c=1$, $\kappa_h=\kappa_c=0.05$, $k_BT_c=0.5$, $k_BT_h=5$. Parameters numerics: $\omega_c=3$, $n=400$, $t=20/\kappa$.}
  \label{fig:non_degen_state_g}
\end{figure}

\begin{figure}[t!]
\centering
\includegraphics[width=\textwidth]{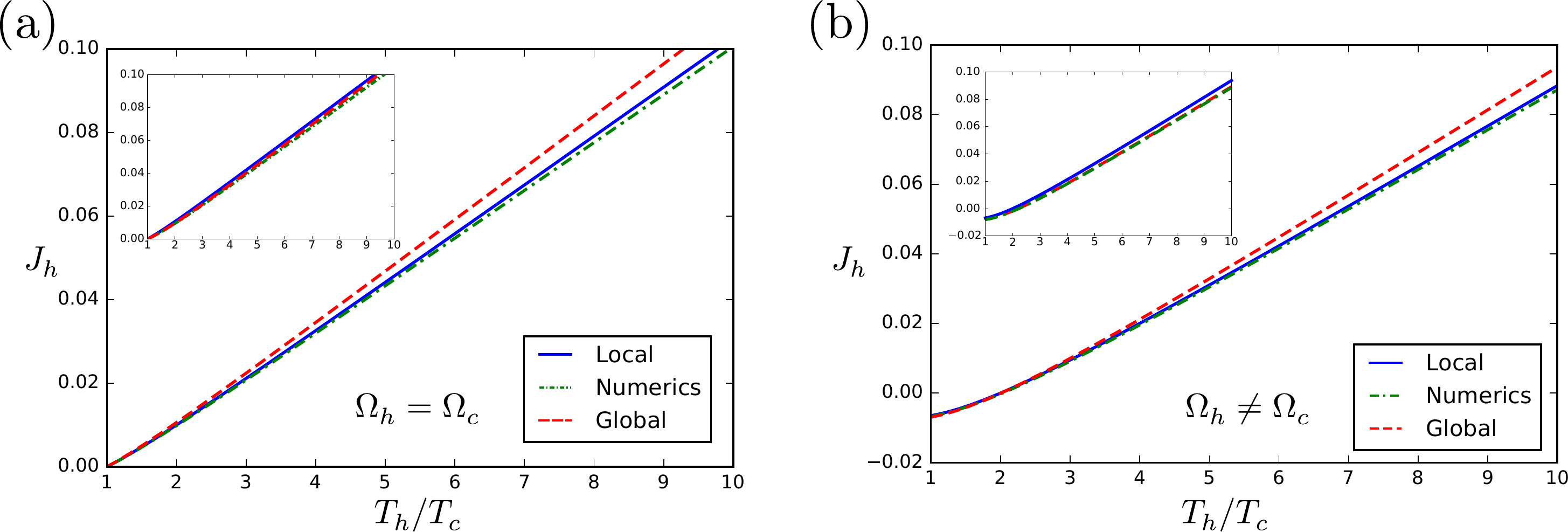}
\caption{Heat currents obtained from the local master equation [cf.~Eq.~\eqref{eq:localmaster}], the global master equation [cf.~Eq.~\eqref{eq:mastersecdegen}], and exact numerics as a function of $T_h$. ${\rm (a)}$ Absence of an external field. ${\rm (b)}$ Presence of an external field. The insets show the heat currents for strong interactions $g=\Omega_c/2$. For all temperatures, the local and the global approach agree well with exact numerics in their respective regimes of validity ($g\ll\Omega_\alpha$ for the local, and $g\gg\kappa_\alpha$ for the global approach).
Parameters: $\Omega_c=1$, $\kappa_h=\kappa_c=0.05$, $g=0.1$ (insets $g=0.5$), $k_BT_c=0.5$, ${\rm (a)}$ $\Omega_h=1$, ${\rm (b)}$ $\Omega_h=2$. Parameters numerics: $\omega_c=3$, $n=400$, $t=20/\kappa$.}
  \label{fig:temperature_heat}
\end{figure}

\subsection{Thermal bias and external field}
Finally, we consider the regime where the considered system performs as a heat engine. This requires both a thermal bias $T_c\neq T_h$ and an external field $\Omega_h\neq\Omega_c$.
In the thermoelectric realization of Ref.~\cite{hofer:2016prb}, this situation corresponds to the presence of a thermal and a voltage bias. In Fig.~\ref{fig:hat_engine_g}, the energy flows through the heat engine are plotted. While the local approach agrees extremely well with exact numerics, the global approach fails for small $g$. Note that all models fulfill $J_c+J_h=P$ (first law), and $\eta<\eta_C$ (second law). For the global approach, we note that it is crucial to use the definitions of energy currents given in Sec.~\ref{sec:globalheat}. If one uses instead definitions similar to the local approach [see Eq.~\eqref{eq:power}], the global approach results in incorrect heat currents, leading in particular to $P=0$ and $J_h = -J_c$. This is consistent with the results of Ref.~\cite{wichterich:2007}, which discusses the absence of any currents as expressed through system observables in the global approach.

\begin{figure}[t!]
\centering
\includegraphics[width=.6\columnwidth]{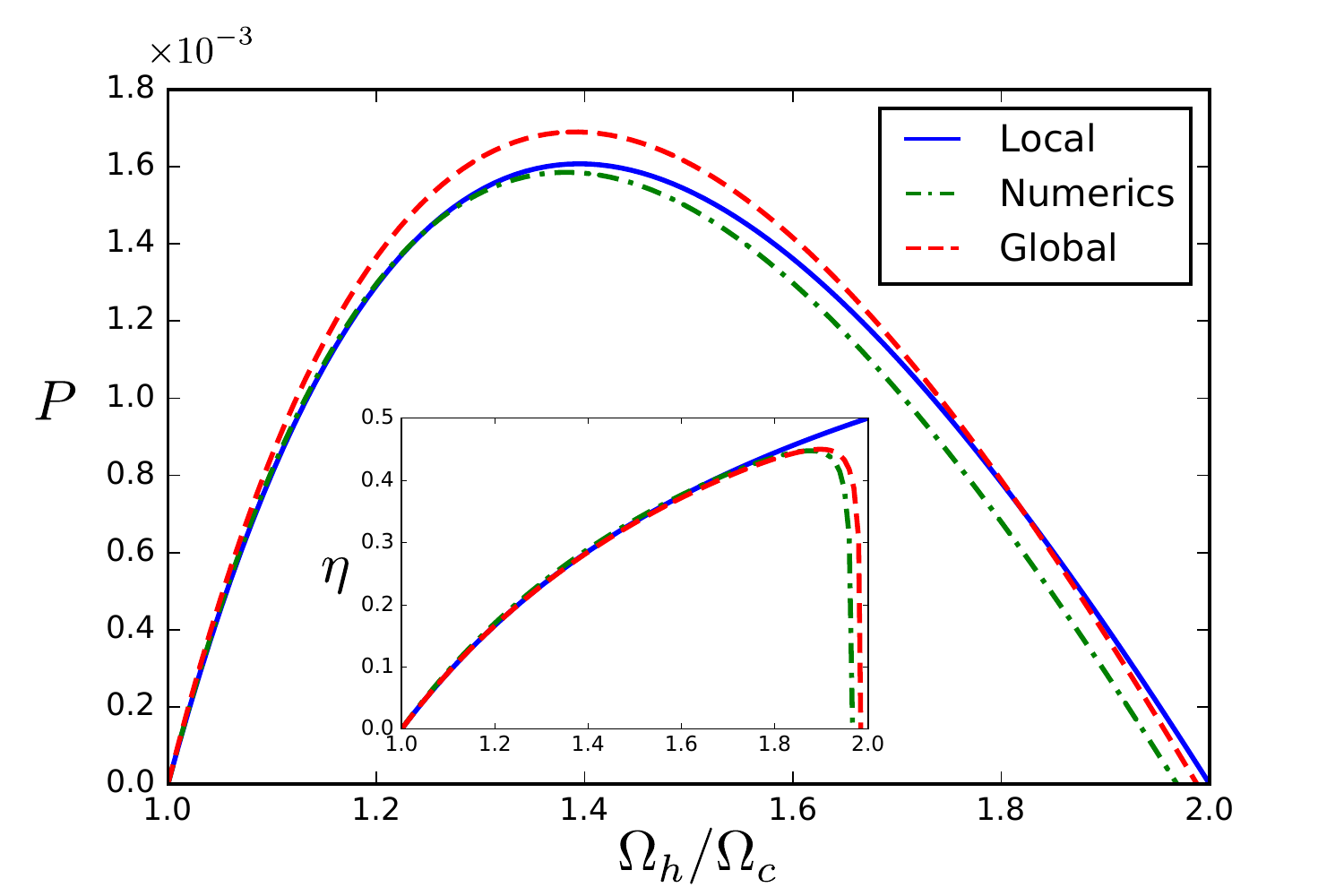}
\caption{Heat engine performance as a function of $\Omega_h$. We find good agreement between the local approach, the global approach, and the exact numerics except for efficiencies close to the Carnot efficiency (here $\eta_C=0.5$). The large differences in efficiencies result from small differences in power and heat currents. 
Parameters: $\Omega_c=1$, $\kappa_h=\kappa_c=0.05$, $g=0.1$, $k_BT_c=0.5$, $k_BT_h=1$. Parameters numerics: $\omega_c=3$, $n=400$, $t=20/\kappa$.}
  \label{fig:power_oh}
\end{figure}

In Fig.~\ref{fig:non_degen_state_g} we compare the steady states against the exact numerics, using again fidelity as a figure of merit. Similarly to the case of a thermal bias without external field, we find that the two approaches give a faithful description in their respective regime of validity.

For completeness, Fig.~\ref{fig:temperature_heat} illustrates the heat currents as a function of temperature in the presence and absence of an external field. These results strengthen the conclusions drawn above: The global approach is valid for $g\gg\kappa_\alpha$ while the local approach is valid for $g\ll\Omega_\alpha$. In particular, the insets show that even at reasonably strong interaction strengths $g$, the local approach agrees well with numerics.

Finally, we further illustrate the heat engine performance in Fig.~\ref{fig:power_oh} which shows the power and efficiency as a function of $\Omega_h$, which determines the external field frequency (given by $\Omega_h-\Omega_c$). We find good agreement in the power as well as the efficiency between the local approach, the global approach, and the exact numerics. Only when the machine is operated close to the Carnot point ($\Omega_h/T_h=\Omega_c/T_c$) do the efficiencies deviate considerably. This is due to the fact that the power and the heat current become very small. Small differences in the energy flows then translate into large differences in the efficiency.

\section{Qubit entangler}
\label{sec:qubits}
\begin{figure}[h!]
\centering
\includegraphics[width=.75\columnwidth]{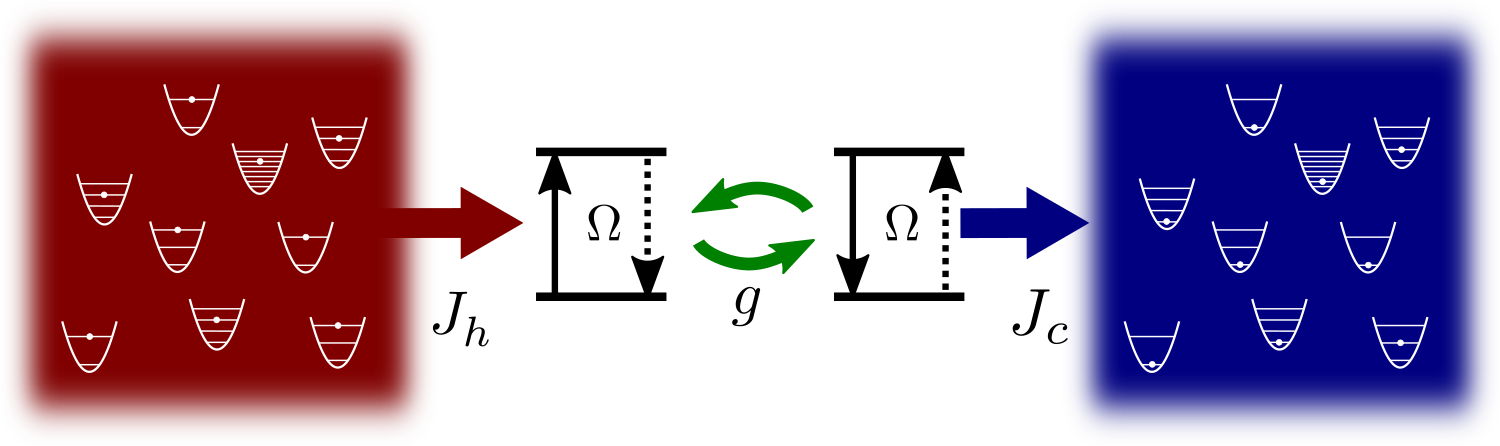}
\caption{Thermal machine consisting of two interacting qubits coupled to bosonic thermal baths.}
  \label{fig:qubitsetting}
\end{figure}

To complete our discussion, we also consider a quantum thermal machine featuring finite-dimensional systems. Specifically we consider a quantum thermal machine consisting of two interacting qubits coupled to separate bosonic thermal baths, as shown in \figref{fig:qubitsetting}, analogous to the setup of \figref{fig:sketch} for harmonic oscillators. This machine can generate entanglement between the two qubits in the steady state, as shown in Ref.~\cite{brask:2015njp}. In the following, however, we focus on comparing the steady states obtained from local and global master equations in the same spirit as above.

We denote the eigenstates of the free Hamiltonians of the qubits by $\ket{0}$, $\ket{1}$, and set the ground state energy to zero. The Hamiltonian of the system is then given by $\hat{H}_{S} = \hat{H}_0 + \hat{H}_{int}$ with
\begin{equation}
\begin{aligned}
\hat{H}_0 & =  \Omega_c \ket{1}\bra{1}\otimes\mathbb{I}_2 + \Omega_h \mathbb{I}_2\otimes\ket{1}\bra{1} \\
\hat{H}_{int} & = g ( \ket{0,1}\bra{1,0} + \ket{1,0}\bra{0,1} ) ,
\end{aligned}
\end{equation}
where $\Omega_c$, $\Omega_h$ are the energy gaps of the qubit, and $g$ is the interaction strength. As in \secref{sec:mastereqs}, we compare local and global master equation models for the evolution of the system. We will focus on the degenerate case where $\Omega_c=\Omega_h=\Omega$.

Both the local and global master equations can be written in Gorini-Kossakowski-Sudarshan-Lindblad (GKSL) form with constant rates \cite{gorini:1976,lindblad:1976}, and so lead to Markovian (specifically semigroup) evolution
\begin{equation}
\partial_t \hat{\rho}(t)  = -i[\hat{H}_S,\hat{\rho}(t)] +
\sum_{\alpha=c,h}\sum_k \Gamma_{\alpha,k} \mathcal{D}[\hat{L}^\dagger_{\alpha,k}] \rho(t) + \bar{\Gamma}_{\alpha,k} \mathcal{D}[\hat{L}_{\alpha,k}] \hat{\rho}(t) ,
\end{equation}
where $\mathcal{D}$ is defined in Eq.~\eqref{eq:lindbladsuper}, $\hat{L}_{\alpha,k}$ are jump operators, and $\Gamma_{\alpha,k}$, $\bar{\Gamma}_{\alpha,k}$ the corresponding rates. We consider bosonic baths for which
\begin{equation}
\Gamma_{\alpha,\varepsilon}  = \kappa_\alpha(\varepsilon) n^\alpha_B(\varepsilon) , \hspace{1cm}
\bar{\Gamma}_{\alpha,\varepsilon}  = \kappa_\alpha(\varepsilon) [n^\alpha_B(\varepsilon) + 1] .
\end{equation}
Here, $\kappa_\alpha(\varepsilon)$ are the bath coupling strengths, and $\varepsilon$ is the (absolute) energy difference associated with the jump induced by $\hat{L}_{\alpha,\varepsilon}$. Thus the $\hat{L}^\dagger_{\alpha,\varepsilon}$ correspond to jumps from lower to higher energies, absorbing energy from the bath $\alpha$, while $\hat{L}_{\alpha,\varepsilon}$ correspond to jumps decreasing the system energy, dissipating energy into the bath $\alpha$.

Local and global master equations for the two-qubit machine can be derived using the same techniques as in \secref{sec:mastereqs}. The system-bath coupling can be taken to have the same form as in Eq.~\eqref{eq:hamiltonian}, with the system annihilation and creation operators replaced by $\hat{A}_c = \hat{\sigma}_-\otimes\mathbb{I}_2$, $\hat{A}_h = \mathbb{I}_2\otimes\hat{\sigma}_-$ and $\hat{A}_c^\dagger = \hat{\sigma}_+\otimes\mathbb{I}_2$, $\hat{A}_h^\dagger = \mathbb{I}_2\otimes\hat{\sigma}_+$ respectively, where $\hat{\sigma}_- = \ket{0}\bra{1}$ and $\hat{\sigma}_+ = \ket{1}\bra{0}$. We take the spectral density of the baths to be Ohmic, as before. The bath coupling strengths are then linear in energy
\begin{equation}
\kappa_\alpha(\varepsilon) = \nu_\alpha \varepsilon ,
\end{equation}
for some constants $\nu_\alpha$. We denote $\kappa_\alpha = \kappa_\alpha(\Omega)$.

For the local master equation, there are just two jump operators, both corresponding to transitions with energy $\Omega$. They are given by 
\begin{equation}
\hat{L}_{c,\Omega}  = A_c = \hat{\sigma}_- \otimes \mathbb{I}_2 ,\hspace{1cm}
\hat{L}_{h,\Omega}  = A_h = \mathbb{I}_2 \otimes \hat{\sigma}_- .
\end{equation}

For the global master equation, the jump operators are found by diagonalizing the system Hamiltonian. Denoting the eigenvalues and eigenstates of $\hat{H}_S$ by $\lambda$ and $\ket{\varphi_\lambda}$ respectively, one has
\begin{equation}
\hat{L}_{\alpha,\varepsilon} = \sum_{\lambda-\lambda'=\varepsilon} \ket{\varphi_{\lambda'}}\bra{\varphi_{\lambda'}} \hat{A}_\alpha \ket{\varphi_\lambda}\bra{\varphi_\lambda} .
\end{equation}
In the degenerate case, $\Omega_c=\Omega_h=\Omega$, the eigenvalues are $0$, $\Omega\pm g$, and $2\Omega$, and the corresponding eigenstates are
\begin{equation}
\ket{\varphi_0}  = \ket{0,0} , \hspace{.75cm}
\ket{\varphi_{\Omega\pm g}}  = (\ket{0,1} \pm \ket{1,0})/\sqrt{2} ,  \hspace{.75cm}
\ket{\varphi_{2\Omega}}  = \ket{1,1} .
\end{equation}
The possible transition energies are $2g$, $\Omega\pm g$, and $2\Omega$. However, only transitions with energies $\Omega\pm g$ can be induced by the system-bath coupling considered here. The non-zero jump operators are
\begin{equation}
\begin{aligned}
\hat{L}_{c,\Omega-g} & = \frac{1}{\sqrt{2}}\ket{\varphi_{\Omega+g}}\bra{\varphi_{2\Omega}} - \frac{1}{\sqrt{2}}\ket{\varphi_0}\bra{\varphi_{\Omega-g}}  , \\
\hat{L}_{c,\Omega+g} & = \frac{1}{\sqrt{2}}\ket{\varphi_{\Omega-g}}\bra{\varphi_{2\Omega}} + \frac{1}{\sqrt{2}}\ket{\varphi_0}\bra{\varphi_{\Omega+g}}  , \\
\hat{L}_{h,\Omega-g} & = \frac{1}{\sqrt{2}}\ket{\varphi_{\Omega+g}}\bra{\varphi_{2\Omega}} + \frac{1}{\sqrt{2}}\ket{\varphi_0}\bra{\varphi_{\Omega-g}}  , \\
\hat{L}_{h,\Omega+g} & = - \frac{1}{\sqrt{2}}\ket{\varphi_{\Omega-g}}\bra{\varphi_{2\Omega}} + \frac{1}{\sqrt{2}}\ket{\varphi_0}\bra{\varphi_{\Omega+g}}  .
\end{aligned}
\end{equation}
We note that these are indeed global in the sense that they involve transitions to and from the non-separable states $\ket{\varphi_{\Omega\pm g}}$.

\begin{figure}[t!]
\centering
\includegraphics[width=\textwidth]{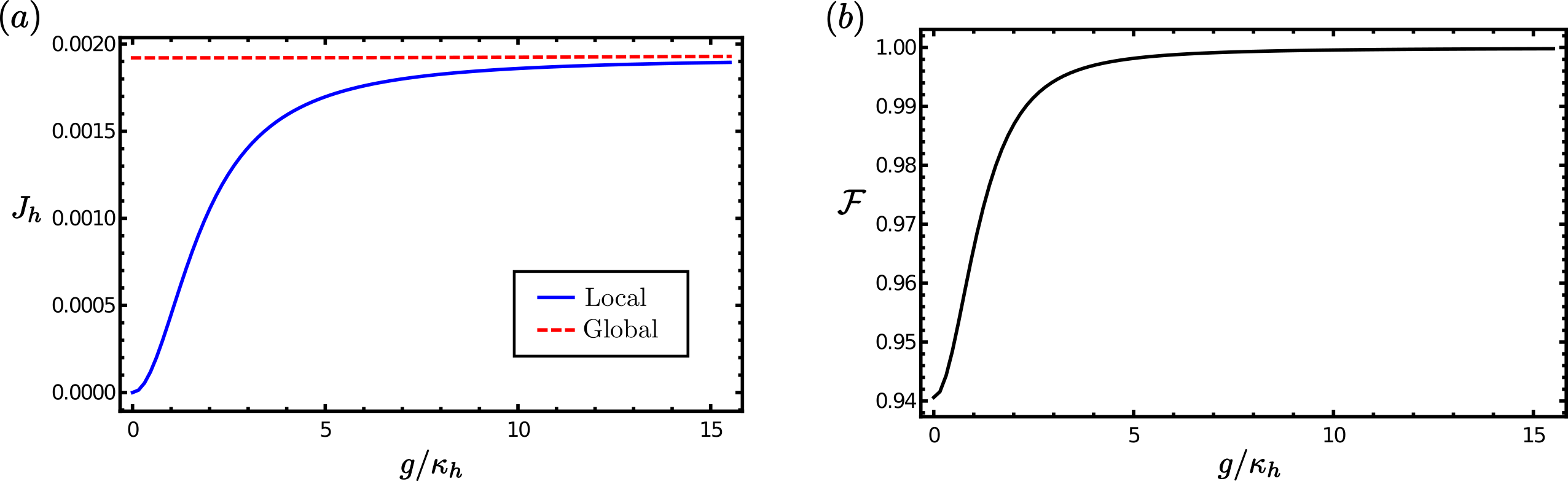}
\caption{\textbf{(a)} Heat current from the hot bath to the system vs the interaction strength in the qubit entangler, for $\Omega=1$, $T_c=0.5$, $T_h=5$, $\kappa_c=0.005$, and $\kappa_h=0.005$. \textbf{(b)} Fidelity between the steady states of the local and global master equations for the same parameters as in (a).}
  \label{fig:qubitFJ}
\end{figure}

We can compute the steady state solutions of both the local and globel qubit master equations. As before, we compare them varying the interaction strength. In \figref{fig:qubitFJ}\,(a) we show the heat current as a function of the interaction strength between the two qubits. As for the harmonic oscillators, the global approach predicts a constant heat current, which is clearly unphysical as $g\rightarrow 0$, while the local model predicts a vanishing heat current in this limit, as expected. The two models agree well for intermediate coupling strength (i.e.~$g/\kappa_\alpha \gtrsim 10$ in this case, note that $\kappa_\alpha$ is taken one order of magnitude smaller than in the previous sections). In \figref{fig:qubitFJ}\,(b) we show the fidelity between the steady states of the two models. Again, we see that the states agree well unless the coupling is weak.

\section{Conclusions}
\label{sec:conclusions}
We investigated the accuracy of the local and global master equations for predicting thermodynamic quantities as well as system steady states describing a quantum heat engine. Exact numerics were used to benchmark the results. We found that the two approaches work very well in their respective regimes of validity which are (for bosonic baths with Ohmic spectral density):
\begin{itemize}
\item Local approach: $g\ll\Omega_\alpha$,
\item Global approach: $g\gg\kappa_\alpha$.
\end{itemize}
More generally, the condition under which the local approach gives a faithful description is given by Eq.~\eqref{eq:justiflocal}. Since the Markov approximation, which underlies both approaches, requires $\kappa_\alpha\ll\Omega_\alpha$, the two regimes of validity overlap. As expected, we find good agreement between the local and the global approach in this region of parameter space.

We note that the local approach is by no means more phenomenological than the global approach. Indeed, the approximation leading to the local master equation is completely analogous to the Markov approximation and has a well defined regime of validity.

Finally, we also investigated a qubit entangler. For this system, no benchmark is available. However, the similarity to the results obtained for the heat engine strongly suggests that similar conclusions with respect to the applicability of the local and the global master equation are valid. We therefore conjecture that our results are qualitatively valid for a variety of baths and system Hamiltonians. As long as the bath-correlation time is much shorter than any inverse inter-system interaction strength, the local approach is valid. The global approach is valid as long as the inter-system interaction strengths are much stronger than the system-bath interaction strengths.

We therefore conclude that the local approach provides a valid description for thermal machines that consist of weakly interacting sub-systems.

\textit{Note added --} During the writing of this manuscript, we became aware of related work \cite{gonzalez:2017}. There the authors also compare local and global master equations, but in the absence of external fields, finding good agreement with the results presented here.

\section*{Acknowledgments}
We acknowledge fruitful discussions with Mark Mitchison and valuable feedback from Ronnie Kosloff, Amikam Levy, Gediminas Kir\v{s}anskas, Andreas Wacker, Felipe Barra, Aashish Clerk, Kay Brandner, and Massimiliano Esposito. M.P.-L. acknowledges support from the
Alexander von Humboldt Foundation. All other authors acknowledge the Swiss National Science Foundation (Starting grant DIAQ, grant $200021\_169002$, Marie-Heim V\"ogtlin grant $164466$, and QSIT). All authors are 
grateful for support from the EU COST Action MP1209 on
Thermodynamics in the quantum regime. 

\appendix

\section{Details on the simulation }
\label{app:numerics}

In this Appendix we discuss the robustness of the numerics, that we take as a benchmark, to the specific choices of $n$ (the number of oscillators) and $\omega_c$ (the cutoff). First of all, recall that we model the bath as a collection of oscillators with frequencies, 
\begin{align}
\label{eq:DiscreteFrequenciesII}
\omega_{k,\alpha}=\frac{k}{n}\omega_{\rm c},
\end{align}
and coupling constants,
\begin{align}
\gamma_{k,\alpha} = \eta_{\alpha} \sqrt{k} \frac{\omega_c}{n}
\label{gamma}
\end{align}
 which models an Ohmic spectral density in the continuous limit. The $\eta_{\alpha}$ in Eq.~\eqref{gamma} are given by $\kappa_{\alpha}=2\pi \eta_{\alpha} \Omega_{\alpha}$.
 
 \begin{figure}[h!]
 \vspace{5mm}
 \centering
 \includegraphics[width=.6\columnwidth]{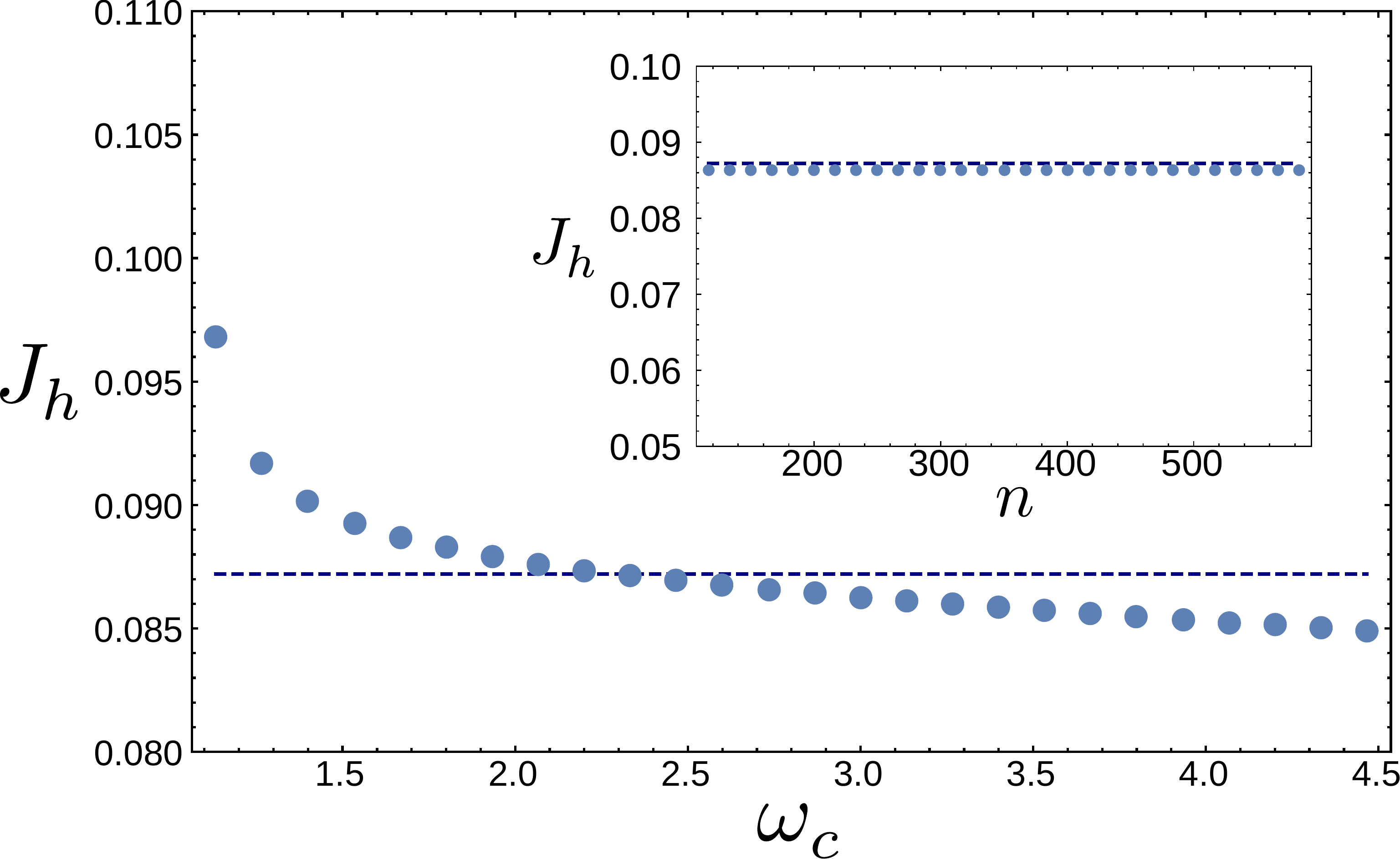}
 \caption{ Dependence of the numerical results on the choice of $n$ and $\omega_c$. Parameters: $\Omega_c=1$, $\Omega_h=1$ $\kappa_h=\kappa_c=0.05$, $\omega_c=3$, $k_BT_c=0.5$, $k_BT_h=5$. Horizontal lines show the predictions of the local master equation in Eq.~\ref{eq:localmaster}.}
   \label{fig:numerics}
 \end{figure}
 
In all the figures of the main text, we take $\omega_c=3$ and $n=400$. In order to see how sensitive our results are to this choice, we plot the heat current as a function of $n$ (the inset) and $\omega_c$, and compare it with the analytic results (using the local approach) in Fig.~\ref{fig:numerics}. It is clearly observed that the results are independent of $n$. On the other hand, we see a small dependence on the values of $\omega_c$. For $\omega_c \approx \Omega_{\alpha}$, the results do not closely match the analytics, which is expected because energetically possible transitions are not captured by the bath. For $\omega_c \geq  2\Omega_{\alpha}$, very good agreement is obtained, with small differences that increase with $\omega_c$. This is due to the renormalization of the Hamiltonian [cf.~Eqs.~\eqref{eq:lambshift} and \eqref{eq:lambshiftdeg}], which is neglected in the analytic calculations. The choice $\omega_c=3$ is hence large enough to capture the different energy transitions, and at the same time not too large so that the effect of the renormalization can be neglected. Similar considerations hold for the case of non-degenerate frequencies of the oscillators of the system. We hence conclude that our numerical benchmark is quite robust to the choice of $n$ and $\omega_c$.

\section*{\refname}

\bibliographystyle{quantum_ph}
\bibliography{biblio}

\end{document}